\pdfoutput=0
\documentclass[journal,12pt,onecolumn]{IEEEtran}

\ifCLASSOPTIONonecolumn
\usepackage{setspace}
\fi
\usepackage[top=2.0cm, bottom=2.2cm, left=2.2cm,right=2.2cm]{geometry}
\IEEEoverridecommandlockouts
\usepackage{cite}
\usepackage{amsmath,amssymb,amsfonts}
\usepackage{adjustbox}
\usepackage{graphicx}
\usepackage{booktabs}
\usepackage{multirow}
\usepackage{placeins}
\usepackage{textcomp}
\usepackage{color}
\usepackage{comment}
\usepackage{psfig}
\usepackage{multirow}
\usepackage{steinmetz} 
\usepackage{tabularx}
\usepackage[multiple]{footmisc} 
\newcommand{\mycomment}[1]{%
}%
\usepackage{environ}
\NewEnviron{nestedcomment}{\mycomment{\BODY}}

\usepackage{algorithm}
\usepackage{algpseudocode}

\usepackage[font=small,labelfont=bf]{caption}

\usepackage{etoolbox}
\makeatletter
\patchcmd{\@makecaption}
  {\scshape}
  {}
  {}
  {}
\makeatother

\def\urltilde{\kern -.15em\lower .7ex\hbox{\~{}}\kern .04em}

\DeclareMathOperator{\sinc}{sinc}

\allowdisplaybreaks

\begin{document}
\title{Received Power Maximization Using Nonuniform Discrete Phase Shifts for RISs With a Limited Phase Range
\thanks{The authors are with the Center for Pervasive Communications and Computing (CPCC), Department of Electrical Engineering and Computer Science,
University of California, Irvine.}
\thanks{This work is partially supported by NSF grant 2030029.}
}

\ifCLASSOPTIONonecolumn
\author{\IEEEauthorblockN{Dogan Kutay Pekcan, {\em Graduate Student Member, IEEE,}}\\
\IEEEauthorblockN{Hongyi Liao, {\em Graduate Student Member, IEEE,}}
\IEEEauthorblockN{\ and Ender Ayanoglu, {\em Fellow, IEEE}}\\
}
\else
\author{\IEEEauthorblockN{Dogan Kutay Pekcan, {\em Graduate Student Member, IEEE}}\\
\IEEEauthorblockN{Hongyi Liao, {\em Graduate Student Member, IEEE}}\\
\IEEEauthorblockN{Ender Ayanoglu, {\em Fellow, IEEE}}\\
}
\fi

\maketitle

\begin{abstract}
To maximize the received power at a user equipment, the problem of optimizing a reconfigurable intelligent surface (RIS) with a limited phase range $R < 2\pi$ and nonuniform discrete phase shifts with adjustable gains is addressed. Necessary and sufficient conditions to achieve this maximization are given. These conditions are employed in two algorithms to achieve the global optimum in linear time for $R \geq \pi$ and $R<\pi$, where $R$ is the limited RIS phase range. With a total number of $N(2K+1)$ complex vector additions, it is shown for $R \geq \pi$ and $R<\pi$ that the global optimality is achieved in $NK$ or fewer and $N(K+1)$ or fewer steps, respectively, where $N$ is the number of RIS elements and $K$ is the number of discrete phase shifts which may be placed nonuniformly
over the limited phase range $R$. In addition, we define two quantization algorithms that we call nonuniform polar quantization (NPQ) algorithm and extended nonuniform polar quantization (ENPQ) algorithm, where
the latter is a novel quantization algorithm for RISs with a significant phase range restriction, i.e., $R<\pi$.
With NPQ, we provide a closed-form solution for the approximation ratio with which an arbitrary set of nonuniform discrete phase shifts can approximate the continuous solution. We also show that with a phase range limitation, equal separation among the nonuniform discrete phase shifts maximizes the normalized performance. Furthermore, we show that the gain of using $K \geq 3$ with $R<\pi/2$ and $K \geq 4$ with $R<\pi$ is only marginal. Finally, we prove that when $R<2\pi/3$, ON/OFF selection for the RIS elements brings significant performance compared to the case when the RIS elements are strictly ON.
\end{abstract}

\section{Problem Definition} \label{sec:signalmodel}
\begin{figure}[!t]
	\centering
    \includegraphics[width=0.30\textwidth]{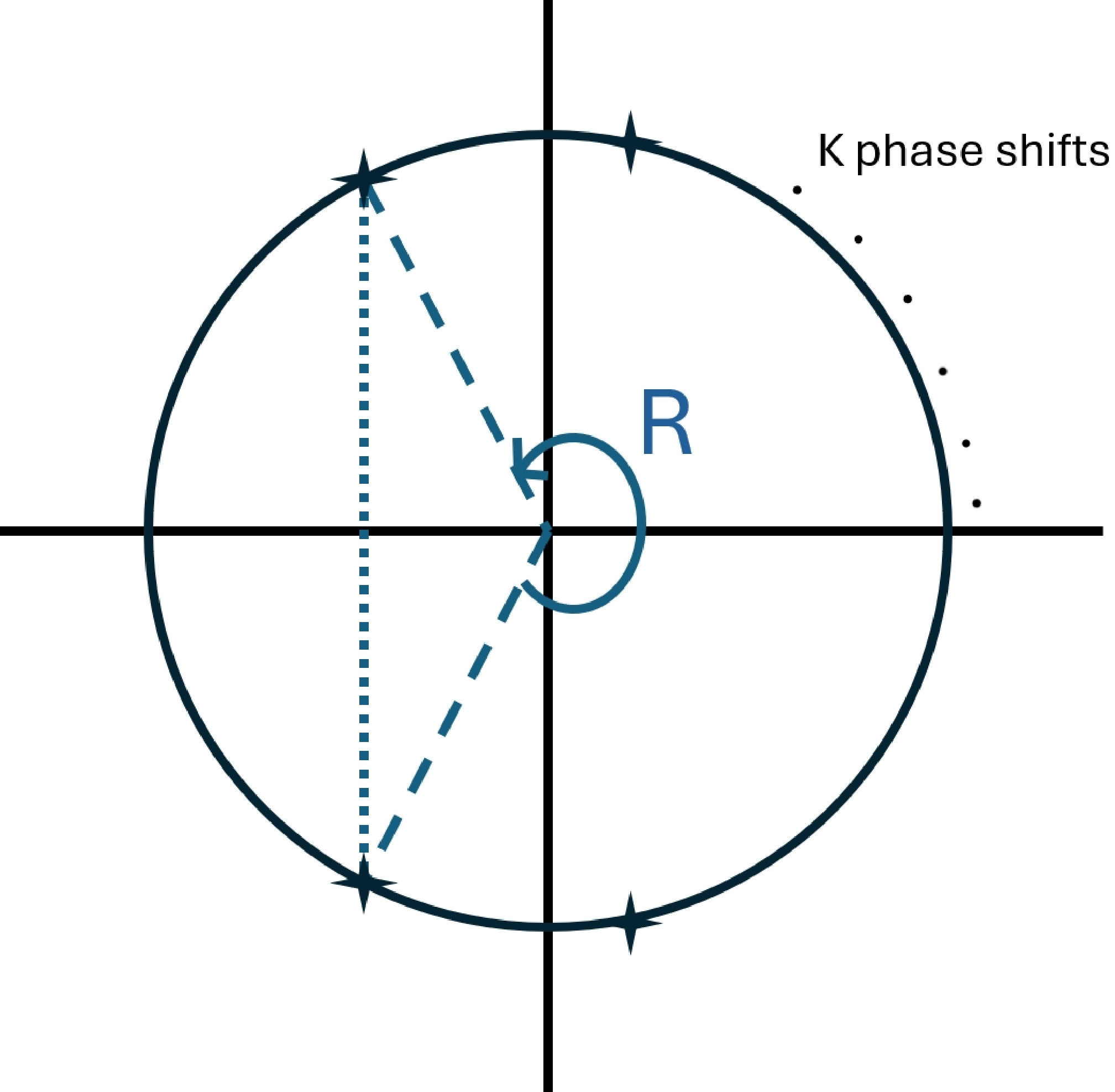}
	\caption{Nonuniform phase placement for $R \in [0,2\pi]$.}
	\label{fig:phaseRange}
\end{figure}

In this paper,
we extend the problem of finding discrete phase shifts to maximize the received power at a User Equipment (UE) for transmission, reflected by a Reflective Intelligent Surface (RIS), originated from a Base Station (BS), see, e.g., \cite{PA24}. In particular,
we address the problem of finding the values $\theta_1, \theta_2, \ldots, \theta_N$ to maximize the received power $\left| h_0 + \sum_{n=1}^N h_n \beta_n^r e^{j\theta_n} \right|^2$, or its square root, where $\theta_n \in \Phi_K$, $\Phi_K = \{\phi_1, \phi_2, \dots, \phi_K\}$ consists of 
discrete phase shifts, $j=\sqrt{-1}$, and $\beta^r_n, \,n=1,\dots,N$ are the RIS gains. With this, the RIS coefficients are given by ${\bf w} = \left[ \beta^r_1 e^{j\theta_1}, \beta^r_2 e^{j\theta_2} \dots, \beta^r_N e^{j\theta_N} \right]$.
We also define the difference among each adjacent phase shift in $\Phi_K$ as $\Omega_K=\{\omega_1, \omega_2, \dots, \omega_K\}$, such that $\phi_{k\oplus1} = \phi_k + \omega_k$\footnote{In this paper, we define $\oplus$ and $\ominus$ to choose from RIS phase shift indexes from $1$ to $K$ as follows.
For $k_1,k_2 \in \{1,\dots,K\}$, $k_1 \oplus k_2 = k_1 + k_2$ if $k_1+k_2 \leq K$ and $k_1 \oplus k_2 = k_1+k_2-K$, otherwise. Similarly, for $k_1,k_2 \in \{1,\dots,K\}$, $k_1 \ominus k_2 = k_1 - k_2$ if $k_1 > k_2$ and $k_1 \ominus k_2 = K + k_1 - k_2$, otherwise.}.
We assume that the main restriction arises due to the RIS phase range $R<2\pi$, therefore the nonuniform phase shifts are selected based on the RIS phase range as in Fig. \ref{fig:phaseRange}.

We will provide optimal and suboptimal but computationally efficient algorithms for the problem. Furthermore, we will analyze the arbitrary phase shift placement and their optimality of approximating the continuous solution for large $N$, in regards to the RIS phase range.

In $\left| h_0 + \sum_{n=1}^N h_n \beta_n^r e^{j\theta_n} \right|$, the values $h_n=\beta_ne^{j\alpha_n}$, $n=0,1,2,\ldots,N$ are the channel coefficients and
$\theta_n,\ n=1,\dots,N$ are the phase values added to the corresponding $h_n$ by an RIS. As for the moment, we let $\beta_n^r=1,\ n=1,\dots,N$, which we will relax after further analysis in this paper.

Initially, the problem can be formally described as
\begin{equation}
\begin{aligned}
 & \underset{\mbox{\boldmath$\theta$}}{\rm maximize\ } f({\mbox{\boldmath$\theta$}})\\
 & {\rm subject\ to\ } \theta_n\in \Phi_K,\ n=1, 2, \ldots, N
\end{aligned}
\label{eqn:eqn1}
\end{equation}
where
\begin{equation}
f({\mbox{\boldmath$\theta$}}) = \bigg|\beta_0e^{j\alpha_0}+\sum_{n=1}^N \beta_n
e^{j(\alpha_n + \theta_n)}\bigg|^2,
\end{equation}
$\beta_n \geq 0,\ n=0,1,\dots,N$, ${\mbox{\boldmath$\theta$}} = (\theta_1, \theta_2, \ldots, \theta_N)$, and $\alpha_n \in [-\pi,\pi)$ for $n=0,1,\dots,N$.

%
\section{Background}
There has been a significant amount of research activity for the selection of discrete phases when the phase range is $2\pi$, see e.g., references in \cite{PA24}. However, the problem of the selection of discrete phases when the phase range is limited to less than $2\pi$ is new and there are only a few works that appeared in the literature or as preprints \cite{HJA23,CYTP,LXMYQ24}. Reference \cite{HJA23} states a uniform phase shift assumption is not realistic in according to the actual behavior of practical RIS elements. The paper maximizes the channel capacity of the target user. It claims to develop a method that finds the optimal reflection amplitudes and phases with complexity linear in the number of RIS elements. Reference~\cite{CYTP} states it models reflection coefficients as discrete complex values that have non-uniform amplitudes and suffer from insufficient phase shift capability. It proposes a group-based query algorithm that takes the imperfect coefficients into consideration. The authors have fabricated an RIS prototype system and validate their theoretical results by experiments. Reference~\cite{LXMYQ24} recognizes that in real-world applications, the phase and bit resolution of RIS units are often non-uniform due to practical requirements and engineering challenges. The authors formulate an optimization problem for discrete non-uniform phase configuration in RIS assisted multiple-input single-output (MISO) communications. They state they propose a partition-and-traversal algorithm which achieves the global optimal solution.

Our work in this paper provides, as an extension of the work in \cite{PA24}, necessary and sufficient conditions for global optimality, two algorithms to achieve the optimum solution which can have smaller number of steps than the works in the literature, and an intuitive quantization algorithm which achieves near-optimal performance with very small complexity. We provide fundamental limits for this approach. We also show that the best solution for this algorithm is obtained when equal separation among the discrete phases in the limited phase range is achieved.

\section{Nonuniform Discrete Phase Shifts and Quantization Solution}\label{ch:quantization}
In this section, we will approach the received power maximization problem with an intuitive quantization algorithm, which we call nonuniform polar quantization (NPQ). This quantization approach is an extension to the uniform polar quantization (UPQ) algorithm proposed in \cite{PA24}\footnote{Note that the terms uniform and nonuniform depend on the range over which they are defined. In this paper, we use the term nonuniform to mean the distribution over the full phase range $[-\pi, \pi)$ is nonuniform, or not equally separated.}. It is similar to the closest point projection (CPP) algorithm in \cite{ZSRLCL22}. Using an analytical approach with this algorithm, we will develop closed-form solutions of the approximation ratios of arbitrary discrete phase shifts to the continuous solution, and develop a framework on how to place the nonuniform discrete phase shifts regarding the RIS phase range.

Consider the problem in (\ref{eqn:eqn1}) but without the condition $\theta_n \in \Phi_K$, $n=1,2,\ldots,N$. We call the solution of this problem the continuous solution to (\ref{eqn:eqn1}).
Given a continuous solution to the problem in (\ref{eqn:eqn1}), say $\theta_n^{\text{cont}}$, NPQ selects the closest possible angle from the set $\Phi_K$. Therefore, for this purpose, we first relax $\theta_n$ and redefine the received power maximization problem as 
\begin{equation}
	\begin{aligned}
		& \underset{\boldsymbol{\theta}^\text{cont}}{\rm maximize\ } f(\boldsymbol{\theta}^\text{cont})\\
		& {\rm subject\ to\ } \theta_n^{\text{cont}} \in [-\pi,\pi),\ n=1, 2, \ldots, N,
	\end{aligned}
	\label{eqn:eqn1rx}
\end{equation}
where
\begin{equation}
f(\boldsymbol{\theta}^\text{cont}) = \bigg|\beta_0e^{j\alpha_0}+\sum_{n=1}^N \beta_n
e^{j(\alpha_n + \theta_n^{\text{cont}})}\bigg|^2.
\end{equation}
In the above equation, $f(\boldsymbol{\theta}^{\text{cont}})$ is calculated by adding $N+1$ complex numbers, where each complex number represents a two-dimensional vector on the complex plane. Among $N+1$ vectors, the only vector we do not have control over is $h_0=\beta_0 e^{j \alpha_0}$. Therefore, in order to achieve the maximum value of $f(\boldsymbol{\theta}^{\text{cont}})$, we can select
\begin{equation}
	\theta_n^{\text{cont}} = \alpha_0 - \alpha_n,\ {\rm for}\  n=1,2,\dots,N,
	\label{eqn:thetaCont}
\end{equation}
so that all vectors will be aligned on top of each other, resulting in the maximum achievable value of $(\sum_{n=0}^{N}\beta_n)^2$. In other words,
\begin{equation}
f(\boldsymbol{\theta}^\text{cont}) = |e^{j\alpha_0}|^2 \bigg|\beta_0+\sum_{n=1}^N \beta_n e^{j(\alpha_n + \theta_n^{\text{cont}}-\alpha_0)}\bigg|^2 = \bigg|\beta_0+\sum_{n=1}^N \beta_n e^{j(\alpha_n + \theta_n^{\text{cont}}-\alpha_0)}\bigg|^2
\end{equation}
and the choice in (\ref{eqn:thetaCont}) maximizes $f(\boldsymbol{\theta}^\text{cont})$.
Given $\theta_n^{\text{cont}} \in [-\pi,\pi)$, NPQ projects to the closest available phase value in $\Phi_K$. Therefore, assuming without loss of generality that $-\pi \leq \phi_1< \phi_2< \dots< \phi_K< \pi$, the decision rule for NPQ is defined as
\begin{equation}
\theta_n^{\text{NPQ}} =
\left\{
\begin{aligned}
\,\,\, \phi_1  & \,\,\,\,\, \text{if}\,\,\,\,\hspace{18mm} -\pi \leq \theta_n^{\text{cont}} < \frac{\phi_1+\phi_2}{2}, \\
\,\,\, \phi_2  & \,\,\,\,\, \text{if}\,\,\,\, \quad \quad \,\,\,\, \frac{\phi_1+\phi_2}{2} \leq \theta_n^{\text{cont}} < \frac{\phi_2+\phi_3}{2}, \\
& \vdots \\
\,\,\, \phi_{K-1} & \,\,\,\,\, \text{if}\,\,\,\, \frac{\phi_{K-2}+\phi_{K-1}}{2} \leq \theta_n^{\text{cont}} < \frac{\phi_{K-1}+\phi_{K}}{2}, \\
\,\,\, \phi_{K} & \,\,\,\,\, \text{otherwise}.
\end{aligned}
\right.
\end{equation}
\normalsize
where $\theta_n^{\text{cont}}$ is the continuous solution in (\ref{eqn:thetaCont}).

From the definition of NPQ, similar to UPQ and CPP approaches, the solution cannot be guaranteed to be globally optimum. In other words, NPQ can only provide a suboptimal solution. Yet, with the quantization approach, the beamforming process can be substantially simplified by using look-up tables, as NPQ only requires $\alpha_n$ for $n=0, 1, \dots, N$ to select the discrete phase shifts.

In the next section, we will analyze the achievable performance under nonuniform discrete phase shift constraints by deriving approximation ratios with NPQ.

\section{Approximation Ratio of Nonuniform Discrete Phase Shifts With NPQ}
Having the quantization approach in hand, we will define an approximation ratio to quantify the effect of the NPQ algorithm, the nonuniform discrete phase shifts, and the RIS phase range on the overall performance of the system. Specifically, the approximation ratio will quantify how well the continuous solution can be approximated. Similar to the approach in \cite{WZ20, PA24}, where we developed an approximation ratio for the uniform polar quantization (UPQ) algorithm \cite{PA24} with uniform discrete phase shifts, we will first approximate the received power $f(\boldsymbol{\theta}^{\text{cont}})$ for large $N$ as 
\begin{align}
	f(\mbox{\boldmath$\theta$}^{\text{NPQ}}) =& \bigg|\beta_0e^{j\alpha_0}+\sum_{n=1}^N \beta_n
	e^{j(\alpha_n + \theta^{\text{NPQ}}_n)}\bigg|^2 \nonumber\\
	= & \left|e^{j\alpha_0}\right|^2 \bigg|\beta_0+\sum_{n=1}^N \beta_n
	e^{j(\alpha_n+\theta^{\text{NPQ}}_n-\alpha_0)}\bigg|^2 \nonumber\\
	= & \bigg|\beta_0+\sum_{n=1}^N \beta_n
	e^{j(\theta^{\text{NPQ}}_n-\theta^{\text{cont}}_n)}\bigg|^2\nonumber\\
	\approx & \bigg|\sum_{n=1}^N \beta_n
	e^{j(\theta^{\text{NPQ}}_n-\theta^{\text{cont}}_n)}\bigg|^2, \label{eqn:frx2aprx}
\end{align}
where the gain from the direct link, i.e., $\beta_0$, is practically discarded for asymptotically large $N$. Let $\delta_n = \theta^{\text{NPQ}}_n-\theta^{\text{cont}}_n$ for $n = 1, 2, \dots, N.$ The resulting absolute square term in (\ref{eqn:frx2aprx}) can be expressed as
\begin{align}
	f({\mbox{\boldmath$\theta$}^{\text{NPQ}}}) &\approx \bigg|\sum_{n=1}^N\beta_n e^{j(\theta^{\text{NPQ}}_n-\theta^{\text{cont}}_n)}\bigg|^2 \nonumber \\
    &= \sum_{n=1}^N \beta_n^2+ \; 2 \sum_{k=2}^N\sum_{l=1}^{k-1} \beta_k\beta_l \cos(\delta_k - \delta_l).
    \label{eqn:frxApxCos}
\end{align}
Assume that in (\ref{eqn:frxApxCos}) all $\beta_k, \beta_l, \delta_k$, and $\delta_l$ are independent from each other. Taking the expectation yields
\begin{equation} \label{eq:Efrx}
	\mathbb{E}[f_{\text{rx}}(\mbox{\boldmath$\theta$}^{\text{NPQ}})] = N\mathbb{E}[\beta_n^2] + N(N-1)\mathbb{E}[\beta_k\beta_l]\mathbb{E}[\cos(\delta_k - \delta_l)].
\end{equation}
Finally, we need to normalize the result in (\ref{eq:Efrx}) with the maximum achievable result to get a ratio from $0$ to $1$, where the continuous solution would achieve $1$. We know from (\ref{eqn:thetaCont}) that the maximum achievable number is $(\sum_{n=0}^{N}\beta_n)^2$. Therefore, $\mathbb{E}[(\sum_{n=0}^{N}\beta_n)^2] = N\mathbb{E}[\beta_n^2] + N(N-1)\mathbb{E}[\beta_k\beta_l]$. As a result, with (\ref{eq:Efrx}) , the ratio of the two expected values can be calculated for asymptotically large $N$ as
\begin{equation}\label{eq:limitENK}
    \lim_{N \to \infty} \frac{\mathbb{E}[f_{\text{rx}}({\mbox{\boldmath$\theta$}^{\text{NPQ}}})]}{\mathbb{E}[(\sum_{n=0}^{N}\beta_n)^2]} = \mathbb{E}[\cos(\delta_k - \delta_l)].
\end{equation}
Hence, $\mathbb{E}[\cos(\delta_k - \delta_l)]$ will be the approximation ratio for NPQ. As we have the independence assumption among $\delta_k$ and $\delta_l$, $\mathbb{E}[\cos(\delta_k - \delta_l)]$ can be simplified further as 
\begin{align}\label{eq:approxCosine}
    \mathbb{E} \left[\cos\left(\delta_k-\delta_l\right)\right] 
    &= \mathbb{E}\left[\cos\left(\delta_k\right)\cos\left(\delta_l\right) + \sin\left(\delta_k\right)\sin\left(\delta_l\right)\right] \nonumber \\
    &= \mathbb{E}\left[\cos\left(\delta_k\right)\cos\left(\delta_l\right)\right] + \mathbb{E}\left[\sin\left(\delta_k\right)\sin\left(\delta_l\right)\right] \nonumber \\
    &= \mathbb{E}\left[\cos\left(\delta_k\right)\right]\mathbb{E}\left[\cos\left(\delta_l\right)\right] + \mathbb{E}\left[\sin\left(\delta_k\right)\right]\mathbb{E}\left[\sin\left(\delta_l\right)\right] \nonumber \\
    &= \left(\mathbb{E}\left[\cos\left(\delta_n\right)\right]\right)^2 + \left(\mathbb{E}\left[\sin\left(\delta_n\right)\right]\right)^2,\ n=1,2,\dots,N.
\end{align}
Therefore, for a given discrete phase shift selection set $\Phi_K$, the approximation ratio can be calculated with (\ref{eq:approxCosine}). We will calculate this for two different scenarios. First, we will provide the approximation ratio for arbitrary $\phi_k,\ k=1,2,\dots, K$, and then for equally separated nonuniform phase shifts over the RIS phase range, as given in Fig. \ref{fig:phaseRange}. In between the two steps, we will also analyze the special connection between the two and show that the latter maximizes the potential of the RIS with nonuniform discrete phase shifts.

\subsection{Arbitrarily Selected Nonuniform Discrete Phase Shifts}
Given the set $\Phi_K$ of arbitrary discrete phase shifts, the approximation ratio will be denoted by $E(\phi_1,\phi_2,\ldots,\phi_K)$. This will be a measure to represent the average performance for an RIS. For this purpose, as a common assumption from the literature to define the quantization error \cite{PA24},\cite{ZSRLCL22},\cite{WZ20}, we will assume that $\theta^{\text{cont}}_n$ is uniformly distributed, i.e., $\theta^{\text{cont}}_n \sim \mathcal{U}[-\pi,\pi]$ to apply the law of total expectation.

Let $\Phi_K = \{\phi_1, \phi_2, \dots, \phi_K\}$ be the set of arbitrarily selected nonuniform phase shifts. Assume without loss of generality that $-\pi \leq \phi_1< \phi_2< \dots< \phi_K<\pi$. Let $\theta^{\text{cont}}_n \in [\phi_k, \phi_{k+1}]$ for $k=1,\ldots,K$ with probability $\frac{\phi_{k+1}-\phi_k}{2\pi}$, in which case $\theta^{\text{NPQ}}_n$ will either be $\phi_k$ or $\phi_{k+1}$. Note that $\delta_n=\theta^{\text{NPQ}}_n -\theta^{\text{cont}}_n$ will be uniformly distributed in $\left[-\frac{\phi_{k+1}-\phi_k}{2},\frac{\phi_{k+1}-\phi_k}{2}\right]$, i.e., $\delta_n \sim \mathcal{U} \left[-\frac{\phi_{k+1}-\phi_k}{2},\frac{\phi_{k+1}-\phi_k}{2}\right]$.

To find $E(\phi_1,\phi_2,\ldots,\phi_K)$, we need to calculate the result in (\ref{eq:approxCosine}). First, note that the distribution of $\delta_n$ is always symmetric around zero, which gives $\left(\mathbb{E}\left[\sin\left(\delta_n\right)\right]\right)^2=0,\, n=1,2,\ldots,N$. Therefore, $E(\phi_1,\phi_2,\ldots,\phi_K) = \left(\mathbb{E}\left[\cos\left(\delta_n\right)\right]\right)^2$. Now, introduce the law of total expectation given as $\mathbb{E}[X] = \mathbb{E}[\mathbb{E}[X|Y]] = \sum_i \mathbb{E}[X|A_i]P(A_i)$, so that $\mathbb{E}\left[\cos\left(\delta_n\right)\right]$ can be calculated as
\begin{align}\label{eq:totalEx_1}
    \mathbb{E} \left[\cos\left(\delta_n\right)\right] = 
    \sum_{k=1}^{K-1} & \left[\frac{\phi_{k+1}-\phi_k}{2\pi}\int_{-(\frac{\phi_{k+1}-\phi_k}{2})}^{(\frac{\phi_{k+1}-\phi_k}{2})} \frac{1}{\phi_{k+1}-\phi_k} \cos(\delta_n) d\delta_n\right] \nonumber \\
    &\hspace{-1em}+\frac{2\pi+\phi_1-\phi_K}{2\pi}\int_{-(\frac{2\pi+\phi_1-\phi_K}{2})}^{(\frac{2\pi+\phi_1-\phi_K}{2})} \frac{1}{2\pi+\phi_1-\phi_K} \cos(\delta_n) d\delta_n
\end{align}
where inside the integral, $\frac{1}{\phi_{k+1}-\phi_k}$ comes from the uniform distribution and $\frac{\phi_{k+1}-\phi_k}{2\pi}$ is the probability of the event $\theta^{\text{cont}}_n \in [\phi_k, \phi_{k+1}]$ occurring. Now, we calculate the term inside the square brackets as
\begin{align}\label{eq:totalExp_integ}
    \frac{\phi_{k+1}-\phi_k}{2\pi} \int_{-(\frac{\phi_{k+1}-\phi_k}{2})}^{(\frac{\phi_{k+1}-\phi_k}{2})} \frac{1}{\phi_{k+1}-\phi_k} \cos(\delta_n) d\delta_n 
    &=\frac{1}{\pi} \int_0^{(\frac{\phi_{k+1}-\phi_k}{2})}\cos(\delta_n) d\delta_n \nonumber \\
    &=\frac{1}{\pi} \sin\left(\frac{\phi_{k+1}-\phi_k}{2}\right).
\end{align}
Similarly, the last term in (\ref{eq:totalEx_1}) will be $\frac{1}{\pi} \sin(\frac{\phi_{K}-\phi_1}{2})$ as $\sin(\frac{2\pi+\phi_{1}-\phi_K}{2})=\sin(\frac{\phi_{K}-\phi_1}{2})$. Therefore, from equations (\ref{eq:approxCosine}), (\ref{eq:totalEx_1}), and (\ref{eq:totalExp_integ}), the approximation ratio for an arbitrary nonuniform discrete phase shift set is
\begin{equation}\label{eq:approxRatio_arbitrary}
    E(\boldsymbol{\phi}) = \frac{1}{\pi^2}\left[\left(\sum_{k=1}^{K-1} \sin\left(\frac{\phi_{k+1}-\phi_k}{2}\right) \right)+ \sin\left(\frac{\phi_{K}-\phi_1}{2}\right)\right]^2,
\end{equation}
where we used the shorthand notation $\boldsymbol{\phi}$ for $\phi_1, \phi_2, \ldots, \phi_K$ with $-\pi \leq \phi_1 < \phi_2 < \cdots < \phi_K < \pi$.
\subsection{How to Place the Nonuniform Discrete Phase Shifts With Limited RIS Phase Range}\label{sec:howtoplace}
In (\ref{eq:approxRatio_arbitrary}), we derived the closed-form expression for the approximation ratio of the set of arbitrary nonuniform discrete phase shifts, i.e., how well the continuous solution can be approximated for large $N$. Now we will prove that given $K$, arranging the phase shifts uniformly will maximize the approximation ratio, and therefore will also maximize the average quantization performance. Define $\Delta_k = (\phi_{k+1}-\phi_k)/{2}$ for $k = 1,2,\ldots, K-1$ and $\Delta_K = (2\pi+\phi_{1}-\phi_K)/{2}$. Note that $\Delta_k \in (0,\pi)$ for $k=1,2,\ldots,K$ and $\sum_{k=1}^K \Delta_k = \pi$. Ignoring the factor 
$1/\pi^2$ in (\ref{eq:approxRatio_arbitrary}), the maximization problem can be equivalently expressed as
\begin{equation}
\begin{aligned}
 & \underset{}{\rm maximize\ } \sum_{k=1}^K \sin(\Delta_k)\\
 & {\rm subject\ to\ } \Delta_1 + \Delta_2 + \cdots + \Delta_K = \pi, \\
 & \quad \quad \quad \quad \,\,\,\, \Delta_k \in (0,\pi),\ k=1, 2, \ldots, K.
\end{aligned}
\label{eq:problemLagrange}
\end{equation}
Using Lagrange multipliers, let
\begin{equation}
    F(\Delta_1,\Delta_2,\ldots,\Delta_K,\lambda) = \sum_{k=1}^K \sin(\Delta_k) + \lambda \left(\sum_{k=1}^K \Delta_k - \pi \right),
\end{equation}
where, the derivatives will be
\begin{align*}
    &\frac{\partial F}{\partial \Delta_k} = \cos(\Delta_k) + \lambda \\
    &\frac{\partial F}{\partial \lambda} = \Delta_1 + \Delta_2 + \cdots + \Delta_K - \pi
\end{align*}
for $k = 1,2,\ldots,K$. Letting $\frac{\partial F}{\partial \Delta_k} = 0$ gives $\cos(\Delta_1) = \cos(\Delta_2) = \cdots = \cos(\Delta_K) = -\lambda$. Since, $\Delta_k \in (0,\pi)$, the solution will be $\Delta_1 = \Delta_2 = \cdots = \Delta_K = \pi/K$ to satisfy the second condition $\frac{\partial F}{\partial \lambda}=0$. Therefore, the optimum placement of the phase shifts is uniformly distributed. Note that this is achievable as long as the RIS phase range $R$ is large enough for a desired number of phase shifts $K$. Therefore, if there is to be a restriction due to the RIS phase range to force nonuniform phase shifts, the condition $R<2\pi\frac{K-1}{K}$ must be satisfied.

At this point, we would like to emphasize an important point regarding the placement of limited phase range $R$ on the unit circle. We remark that the symmetry between the phase shifts $-\frac{R}{2}$ and $\frac{R}{2}$ in Fig. \ref{fig:phaseRange} is not 
required and the techniques presented in this paper are applicable to any nonuniform discrete phase shifts structure
with a total contiguous phase range $R$. 
Because, for an arbitrary nonuniform phase shift structure, the RIS phase range would satisfy the condition $R = 2\pi - \omega_{\Bar{k}}$ where $\omega_{\Bar{k}}$ is the largest value in the set $\Omega_K$.
So, without loss of generality, we will use the approach in Fig. \ref{fig:phaseRange}, i.e., $-\pi \leq \phi_1 < \cdots < \phi_K = \phi_1+R < \pi$ with $R<2\pi\frac{K-1}{K}$.
The condition $R<2\pi\frac{K-1}{K}$ arises due to the fact that $R$ comes from $\omega_{\Bar{k}}$, and the condition $\omega_{\Bar{k}} \geq \frac{2\pi}{K}$ must be satisfied as $\sum_{k=1}^K \omega_k = 2\pi$ and by its definition $\omega_{\Bar{k}} \geq \omega_k,$ for $k \in \{1,2,\ldots,K\} \setminus \Bar{k}$.
Note that, this will make sure that the discrete phase shifts cannot be placed uniformly over the unit circle. In addition, while we recognize the phase range is not necessarily symmetric, we will assume the discrete phase shifts to be distributed over the range $[-\frac{R}{2},\frac{R}{2}]$ without loss of generality.
\subsubsection{Extension of Analysis With RIS Phase Range Restriction}
When there is a sufficient restriction due to the RIS, i.e., $R<2\pi\frac{K-1}{K}$, there is no way that the arbitrary discrete phase shifts can be distributed uniformly over the range $[-\pi,\pi)$. However, we can still question the placement of the discrete phase shifts over the range $R$ that the RIS can reach and show that equally separated discrete phase shifts over the range $R$ will maximize the performance. Now, without loss of generality, let $-\pi \leq \phi_1 < \cdots < \phi_K = \phi_1+R < \pi$ with $R<2\pi\frac{K-1}{K}$, as given in Fig. \ref{fig:phaseRange}. Substituting $\phi_K = \phi_1+R$ in (\ref{eq:approxRatio_arbitrary}), we have
\begin{equation}\label{eq:approxRatio_temp}
E(\boldsymbol{\phi}) = \frac{1}{\pi^2}\left[\bigg(\sum_{k=1}^{K-2}
\sin\left(\frac{\phi_{k+1}-\phi_k}{2}\right)\bigg) +
\sin\left(\frac{\phi_{1}+R-\phi_{K-1}}{2}\right)+ \sin\left(\frac{R}{2}\right)\right]^2,
\end{equation}
where it is clear that $R$ will directly impact the average performance, 
we leave the discussion of this to the next section. Now, focusing on the placement of discrete phase shifts, we will omit the $\sin\left(\frac{R}{2}\right)$ term. Note from (\ref{eq:approxRatio_temp}) that this time we need to define $\Delta_k'$ for $k' = 1,\ldots,K-1$. Therefore, let $\Delta_{k'} = \frac{\phi_{k'+1}-\phi_{k'}}{2},\ k'=1,\ldots,K-2$ and $\Delta_{K-1}=\frac{\phi_1+R-\phi_{K-1}}{2}$. Similar to the arbitrary case, using the Lagrange multipliers, we define the equivalent maximization problem as
\begin{equation}
\begin{aligned}
 & \underset{}{\rm maximize\ } \sum_{k'=1}^{K-1} \sin(\Delta_{k'})\\
 & {\rm subject\ to\ } \Delta_1 + \Delta_2 + \cdots + \Delta_{K-1} = \frac{R}{2}, \\
 & \quad \quad \quad \quad \,\,\,\, \Delta_k \in \left(0,\frac{R}{2}\right),\ k'=1, 2, \ldots, K-1.
\end{aligned}
\label{eq:problemLagrange2}
\end{equation}
Define
\begin{equation}
    F'(\Delta_1,\Delta_2,\ldots,\Delta_{K-1},\lambda) = \sum_{k'=1}^{K-1} \sin(\Delta_{k'}) + \lambda \left(\sum_{k'=1}^{K-1} \Delta_{k'} - \frac{R}{2} \right),
\end{equation}
where, the derivatives will be
\begin{align*}
    &\frac{\partial F}{\partial \Delta_{k'}} = \cos(\Delta_{k'}) + \lambda \\
    &\frac{\partial F}{\partial \lambda} = \Delta_1 + \Delta_2 + \cdots + \Delta_{K-1} - \frac{R}{2}
\end{align*}
for $k' = 1,\ldots,K-1$. Letting $\frac{\partial F}{\partial \Delta_{k'}} = 0$ gives $\cos(\Delta_1) = \cos(\Delta_2) = \cdots = \cos(\Delta_{K-1}) = -\lambda$. Since $\Delta_{k'} \in (0,\frac{R}{2})$ and in this range the cosine function is monotonically decreasing, the solution is provided by $\Delta_1 = \Delta_2 = \cdots = \Delta_{K-1} = \frac{R}{2(K-1)}$. Note that this also satisfies $\frac{\partial F}{\partial \lambda}=0$. Therefore, the optimum placement of the phase shifts is equally separated over the range $R$ to maximize the average normalized performance of the RIS.

In the following section, we derive the approximation ratio for the nonuniform phase shifts with the RIS phase range restriction, i.e., $R<2\pi\frac{K-1}{K}$, where the discrete phase shifts are equally separated. This placement of the nonuniform phase shifts will also be adopted for the rest of the paper, including the numerical results, as suggested by the performance maximization approach and practicality.
\subsection{Practically Selected Nonuniform Discrete Phase Shifts}
We have shown that given the RIS phase range $R$, the placement of the nonuniform discrete phase shifts over the RIS phase range needs to be equally separated, to harness the potential of the RIS and maximize the approximation ratio. Therefore, in Fig. \ref{fig:phaseRange}, we let $\Phi_K = \left\{-\frac{R}{2}, \frac{R}{K-1}-\frac{R}{2}, 2\frac{R}{K-1}-\frac{R}{2},\ldots,\right.$ $\left. (K-1)\frac{R}{K-1}-\frac{R}{2}\right\}$. So that, with the equally separated discrete phase shifts, the decision rule for the NPQ can alternatively be defined as
\begin{equation}\label{eq:thetaNPQ}
\theta_n^{\text{NPQ}} =
\left\{
\begin{array}{lll}
\frac{R}{2} & \text{if} & \frac{R}{2} \leq \theta_n^{\text{cont}},\\
\bigg\lfloor \frac{\theta_n^{\text{cont}}+\frac{R}{2}}{\omega'} \bigg\rceil \times \omega' - \frac{R}{2} & \text{if} & -\frac{R}{2} \leq \theta_n^{\text{cont}} < \frac{R}{2}, \\
-\frac{R}{2} & \text{if}  & \theta_n^{\text{cont}} < -\frac{R}{2},
\end{array}
\right.
\end{equation}
where $\lfloor \cdot \rceil$ is the rounding function defined as $\lfloor x \rceil = {\rm sgn} (x)\left\lfloor |x| + 0.5 \right\rfloor.$

\begin{figure}[!t]
	\centering
    \includegraphics[width=0.45\textwidth]{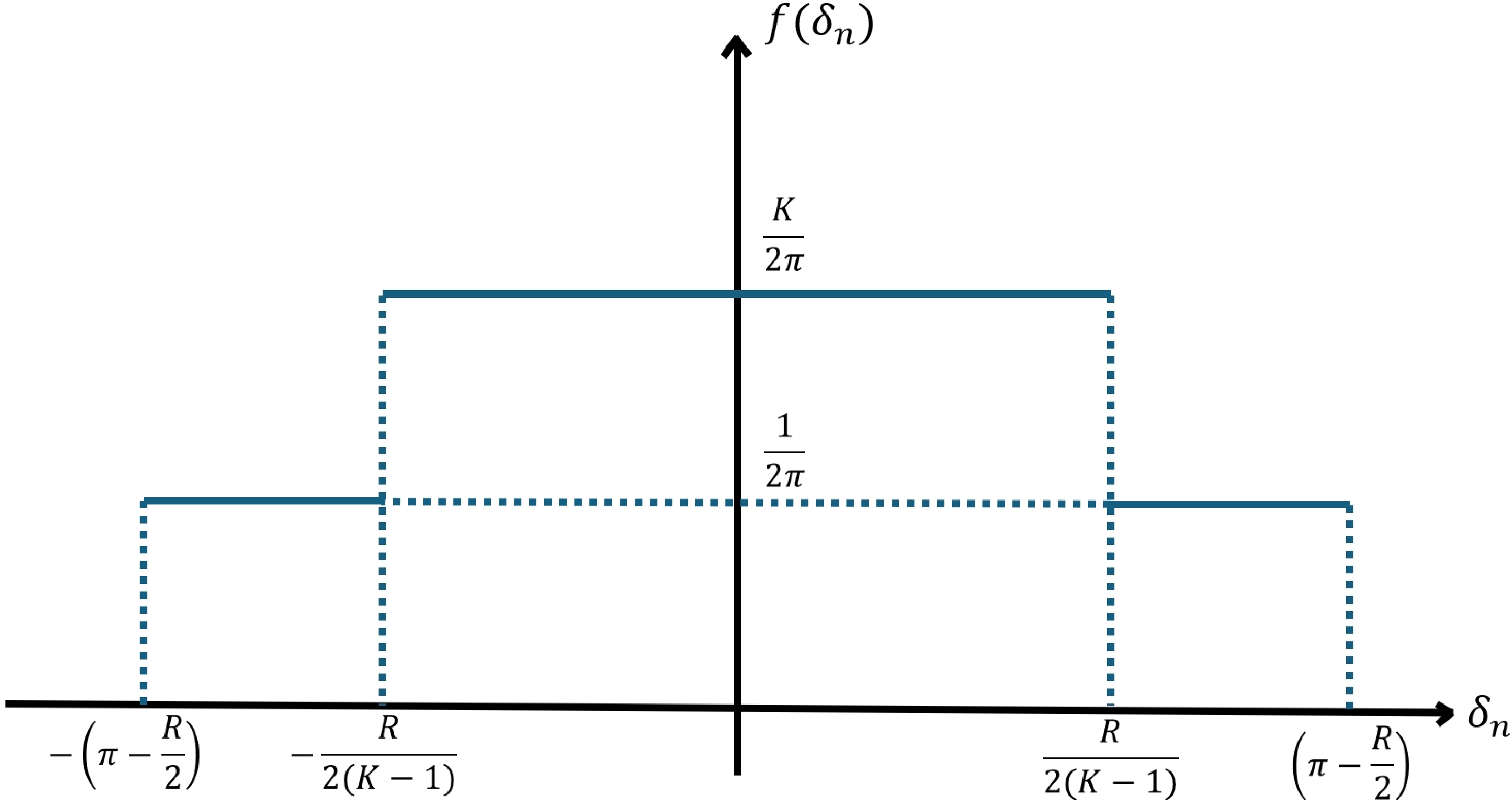}
	\caption{PDF of $\delta_n$, i.e., the quantization error.}
	\label{fig:PDFdeltan}
\end{figure}

Let us define the approximation ratio as $E(R,K) = \mathbb{E}\left[\cos\left(\delta_k-\delta_l\right)\right]$, where we have $\delta_n = \theta^{\text{NPQ}}_n-\theta^{\text{cont}}_n$.
From the definition of $\theta^{\text{NPQ}}_n$ and $\theta^{\text{cont}}_n$ in (\ref{eq:thetaNPQ}) and (\ref{eqn:thetaCont}) respectively, clearly $\delta_n \in \left[-(\pi-\frac{R}{2}), \pi-\frac{R}{2}\right]$.
Remembering the assumption that $\theta^{\text{cont}}_n \sim \mathcal{U}[-\pi,\pi]$, the probability density function (PDF) of $\delta_n$, i.e., $f(\delta_n)$, can be deduced simply and it is plotted in Fig. \ref{fig:PDFdeltan}.
\begin{figure}[!t]
	\centering
\includegraphics[width=0.48\textwidth]{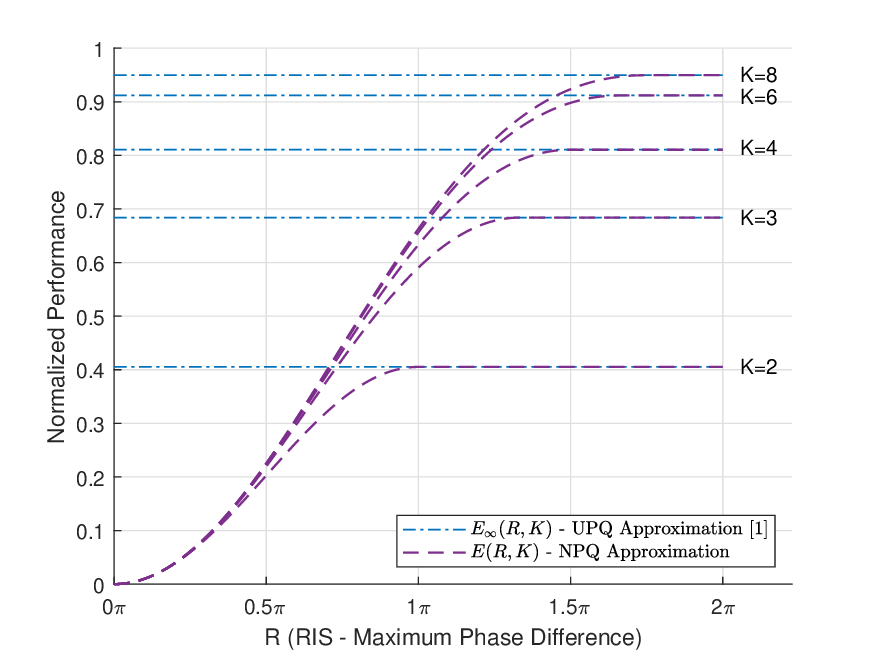}
	\caption{$E(R,K)$ vs $R$ for $K \in \{2,3,4,6,8\}$}
	\label{fig:ERKtheoreticPlot}
\end{figure}
With the PDF $f(\delta_n)$, we need to calculate the simplified version of the term $\mathbb{E}\left[\cos\left(\delta_k-\delta_l\right)\right]$ as given in (\ref{eq:approxCosine}). Note that, the second term in (\ref{eq:approxCosine}) will be zero, since $f(\delta_n)$ is an even function. Therefore, we only need to calculate $\left(\mathbb{E}\left[\cos\left(\delta_n\right)\right]\right)^2$ to find $E(R,K)$. Let us first calculate $\mathbb{E}[\cos(\delta_n)]$ as
\begin{align}
\mathbb{E} \left[\cos\left(\delta_n\right)\right] 
    &= 2 \left[\int_{0}^{\frac{R}{2(K-1)}} \cos(\delta_n) \frac{K}{2\pi}\,d\delta_n + \int_{\frac{R}{2(K-1)}}^{\pi-R/2} \cos(\delta_n) \frac{1}{2\pi}\,d\delta_n\right] \nonumber \\
    &= \frac{1}{\pi} \bigg[K\left[\sin\left(\frac{R}{2(K-1)}\right)-\sin(0)\right] +\left[\sin\left(\pi-\frac{R}{2}\right) - \sin\left(\frac{R}{2(K-1)}\right) \right] \bigg] \nonumber \\
    &= \frac{1}{\pi}\left[(K-1)\sin\left(\frac{R}{2(K-1)}\right) + \sin\left(\frac{R}{2}\right)\right] \label{eq:approxSin}\\
    &= \frac{R}{2\pi}\left[{\rm sincu}\left(\frac{R}{2(K-1)}\right) + {\rm sincu}\left(\frac{R}{2}\right)\right] \label{eq:approxSincu}
\end{align}
where from (\ref{eq:approxSin}) to (\ref{eq:approxSincu}), we divide and multiply by $R/2$, and ${\rm sincu}(\cdot)$ represents the unnormalized sinc function ${\rm sincu}(x) = \frac{\sin x}{x}$. Note that ${\rm sincu}(x) = \sinc(\frac{x}{\pi})$ Also, note that (\ref{eq:approxSin}) is compatible with (\ref{eq:approxRatio_arbitrary}) with $\frac{\phi_{k+1}-\phi_k}{2} = \frac{R}{2(K-1)}$ for $k=1,\ldots,K-1$ and $\frac{\phi_K-\phi_1}{2} = \frac{R}{2}$. Thus, the approximation ratio for the NPQ algorithm is
\begin{equation}
    E(R,K) = \frac{R^2}{4\pi^2}\left[\sinc\left(\frac{R}{2\pi(K-1)}\right) + \sinc\left(\frac{R}{2\pi}\right)\right]^2
\end{equation}
where $R$ is the RIS phase range and $\sinc(\cdot)$ is normalized satisfying $\sinc(1)=0$.
An illustration for the theoretical calculations of the approximation $E(R,K)$ is given in Fig.~\ref{fig:ERKtheoreticPlot}, where it can be seen that $E(R,K)$ converges to the approximation ratio of the uniform phases, i.e., $E_{\infty}(K)$ in \cite{PA24}, as the RIS phase range increases.
From our analysis of the optimum selection of nonuniform discrete phases in Section~\ref{sec:howtoplace}, we know that the equal separation in the RIS phase range will maximize the average performance.
Even with the best case scenario with the optimal placement of the nonuniform phases, Fig. \ref{fig:ERKtheoreticPlot} shows that the gain of using $K\geq3$ is only marginal when $R<\pi/2$.
Similarly, the gain of using $K=4$ or more discrete phase shifts is negligible unless the RIS phase range is large enough, i.e., $R>\pi$.
We remark that the approximation ratio is calculated for sufficiently large $N$. Further analysis to confirm the validity of the theoretical calculation of $E(R,K)$ is provided in the numerical results.

In the next section, we will define our nonuniform discrete phase shift selection algorithm that guarantees the global optimal solution for $\beta_n^r=1,\ n=1,\ldots,N$, or equivalently when $R\geq \pi$, and it will be an extension of \cite[Algorithm~1]{PA24}. We further improve it in the sequel by relaxing $\beta_n^r$ in the interval $[0,1]$ to improve the performance whenever the RIS phase range is less than $\pi$, i.e., $R<\pi$.

\section{Optimal Solution With Nonuniform Discrete Phase Shifts}\label{ch:globalOptimum}
In this section, we aim to solve the received power maximization problem, so that we can get the global optimum solution in linear time. We want to maximize $\left| h_0 + \sum_{n=1}^N h_n e^{j\theta_n} \right|$ where $h_n = \beta_ne^{j\alpha_n}$ for $n = 0, 1, \ldots, N$, and ${\mbox{\boldmath$\theta$}} = (\theta_1, \theta_2, \ldots, \theta_N)$. Define $g$ as
\begin{equation}
g = h_0 + \sum_{n=1}^N h_n e^{j\theta_n^*}\label{eqn:gstar}
\end{equation}
where $\theta^*_n$ are the discrete phase shifts that lead to the global optimum. Let $\mu=g/|g|$ so that $|g| = g e^{(-j\phase{\mu})}$. Similar to the condition in \cite{PA24}, we can make use of the following lemma.

{\em Lemma~1:\/} For an optimal solution $(\theta_1^*, \theta_2^*, \ldots,
\theta_n^*)$, it is necessary and sufficient that each $\theta_n^*$ satisfy
\begin{equation}
\theta_n^* = \arg \max_{\theta_n\in \Phi_K} \cos(\theta_n + \alpha_n -\phase{\mu})
\label{eqn:lemma2}
\end{equation}
for an arbitrary $\Phi_K$.

{\em Proof:\/} We can rewrite $|g| = g e^{(-j\phase{\mu})}$ as
\begin{align}
|g| =& \ \beta_0 e^{j(\alpha_0-\phase{\mu})} + \sum_{n=1}^N \beta_n e^{j(\alpha_n+\theta_n^*-\phase{\mu})} \nonumber\\
 = & \ \beta_0 \cos(\alpha_0 - \phase{\mu}) + j \beta_0 \sin (\alpha_0-\phase{\mu}) + \sum_{n=1}^N \beta_n \cos(\theta_n^* + \alpha_n - \phase{\mu}) 
+ j \sum_{n=1}^N \beta_n \sin(\theta_n^* + \alpha_n - \phase{\mu}).
\label{eqn:absg}
\end{align}
Because $|g|$ is real-valued, the second and fourth terms in (\ref{eqn:absg}) sum to zero, and
\begin{equation}\label{eq:resulting_g}
|g| = \beta_0 \cos(\alpha_0 - \phase{\mu}) + \sum_{n=1}^N \beta_n \cos(\theta_n^* + \alpha_n - \phase{\mu}),
\end{equation}
from which (\ref{eqn:lemma2}) follows as a necessary and sufficient condition for the lemma to hold.
\hfill$\blacksquare$

With the help of this lemma, we have the necessary and sufficient conditions to get the optimal phase shift selections.
However, at this point, we assumed that the optimum $\mu$ would be given.
To make use of this mathematical conditioning on the globally optimum solution, we need an operational framework to find $\mu$, similar to \cite{PA24,b1}.
While $\mu$ can be anywhere on the unit circle, given the channel realizations $h_n$ for $n=0,1,\dots,N$, we provide the following proposition to reduce the search space of $\mu$ to a finite size, as an extension to \cite[Proposition~1]{PA24}.
Towards that end, we will define the following sequence of complex numbers with respect to each $n=1,2,\ldots,N$ as
\begin{equation} \label{eq:snk_new}
    s_{nk} = \exp\left({j\left(\alpha_n + \phi_k - \frac{\omega_{k \ominus 1}}{2}\right)}\right),\ {\rm for}\ k=1,2,\ldots,K.
\end{equation}
Define, for any two points $a$ and $b$ on the unit circle $C$, ${\rm arc}(a:b)$ to be the unit circular arc with $a$ as the initial end and $b$ as the terminal end in the counterclockwise direction, with the two endpoints $a$ and $b$ being excluded.

{\em Proposition 1:\/} A sufficient condition for $\theta_n^*=\phi_k$ is
\begin{equation}
\mu \in {\rm arc} (s_{nk}:s_{n,k+1}).
\label{eqn:prop1}
\end{equation}

\begin{figure}[!t]
	\centering
    \includegraphics[width=0.45\textwidth]{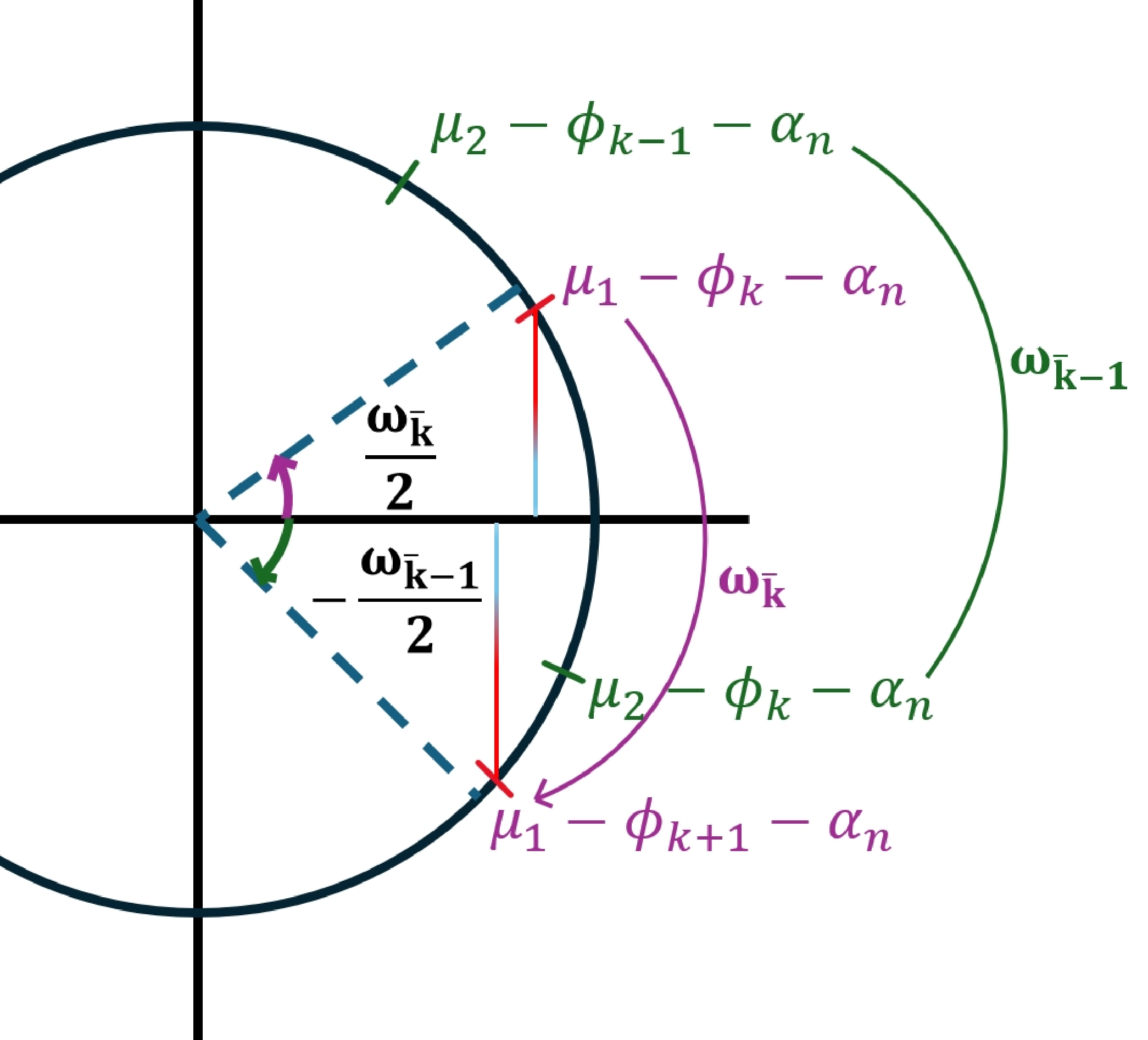}
	\caption{An illustration for the optimality of $\theta_n^*=\phi_k$ given $\mu \in {\rm arc} (s_{nk}:s_{n,k+1})$.}
	\label{fig:prop1}
\end{figure}

{\em Proof:\/} Assume $\mu$ satisfies (\ref{eqn:prop1}). Then,
\begin{equation}
    \phase\mu \in \left(\alpha_n + \phi_k - \frac{\omega_{k-1}}{2}, \alpha_n + \phi_{k+1} - \frac{\omega_{k}}{2}\right).
\end{equation}
By subtracting $\theta_n$ and $\alpha_n$, we get
\begin{equation}
    \phase\mu-\theta_n-\alpha_n \in \left(\phi_k - \frac{\omega_{k-1}}{2}-\theta_n, \phi_{k+1} - \frac{\omega_{k}}{2}-\theta_n\right).
\end{equation}
Now, let $\theta_n = \phi_k$. Then,
\begin{equation}
    \phase\mu-\theta_n-\alpha_n \in \left(\phi_k - \frac{\omega_{k-1}}{2} - \phi_k, \phi_{k+1} - \frac{\omega_{k}}{2} - \phi_k\right).
\end{equation}
By substituting $\phi_{k+1} = \phi_k + \omega_k$, we have
\begin{equation}\label{eq:insideCos}
    \phase\mu-\theta_n-\alpha_n \in \left(-\frac{\omega_{k-1}}{2}, \frac{\omega_{k}}{2}\right).
\end{equation}
Therefore, letting $\theta_n = \phi_k$ results in the largest $\cos(\theta_n + \alpha_n - \phase\mu)$ value among other possibilities for $\mu$, as illustrated in Fig.~\ref{fig:prop1} by showing the effect of selecting the phase shift option before and after than $\phi_k$. Since $\cos(\phase\mu - \theta_n - \alpha_n) = \cos(\theta_n + \alpha_n - \phase\mu)$, the proof is complete.
\hfill$\blacksquare$

Finally, to operate with Proposition~1, we will eliminate duplicates among $s_{nk}$ and sort to get $e^{j\lambda_l}$ such that $0\le \lambda_1 < \lambda_2 < \cdots < \lambda_L < 2\pi.$ Define the update rule as
\begin{equation}\label{eq:updateRule}
    {\cal N} (\lambda_l) = \{ \{n',k'\} | \phase{s_{n'k'}} = \lambda_l \}.
\end{equation}
Let us search for the optimum $\mu$ by traversing the unit circle in the counterclockwise direction, starting from $\phase{\mu}=0$. With Proposition~1, we know that $\theta_n$ for $n=1, 2, \ldots, N$ will remain the same unless $\mu$ switches from one arc to another. Whenever $\mu$ switches arcs, there exists $n$ such that $\theta_n$ will be updated, i.e., if
\begin{equation}
\mu\in {\rm arc}{(e^{j\lambda_l}:e^{j\lambda_{l+1}})} \rightarrow \mu\in {\rm arc}(e^{j\lambda_{l+1}}:e^{j\lambda_{l+2}}),
\label{eqn:eqn40}
\end{equation}
then for every $\{n',k'\}$, $\theta_{n'}$ must be updated according to the update rule in (\ref{eq:updateRule}) as
\begin{equation}
\theta_{n'} \rightarrow \phi_{k'}, \quad \{n',k'\} \in {\cal N}(\lambda_{l+1}).
\label{eqn:eqn41}
\end{equation}
Therefore, the optimum solution will come from $L \leq NK$ possible candidates of $\mu$. For each candidate, we will operate using the sufficiency condition in Proposition~1 that is guaranteed to provide the globally optimum solution, since it is compatible with {\em Lemma~1}.

We now specify Algorithm~1 as the generalized version of \cite[Algorithm~1]{PA24} to work with non-uniform phase shifts and achieve the global optimum in $L \leq NK$ steps. We remark that, for uniformly distributed phase shifts, we showed in \cite{PA24} that the convergence can be achieved in $N$ or \textit{fewer} steps.
\begin{figure*}[!t]
\centering
\begin{minipage}{0.48\textwidth}
\centering
\includegraphics[width=1.1\textwidth]{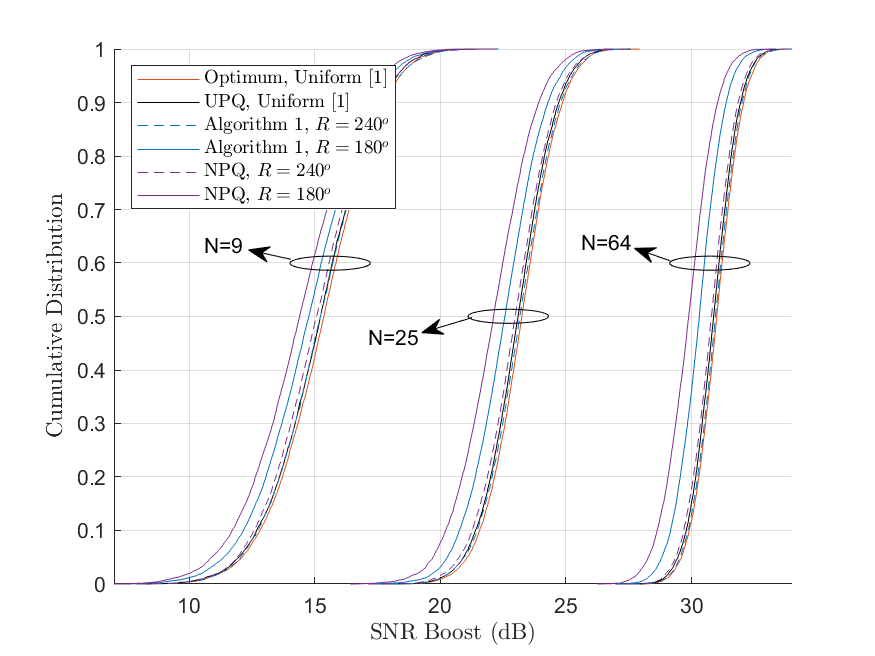}
\caption{CDF plots for SNR Boost with \cite[UPQ]{PA24}, \cite[Algorithm~1]{PA24}, Algorithm~1, and nonuniform polar quantization (NPQ), for $K=4$.}
 \label{fig:SNRBoost_CDF_K4}
\end{minipage}%
\hspace{0.03\textwidth}
\begin{minipage}{0.48\textwidth}
\centering
\includegraphics[width=1.1\textwidth]{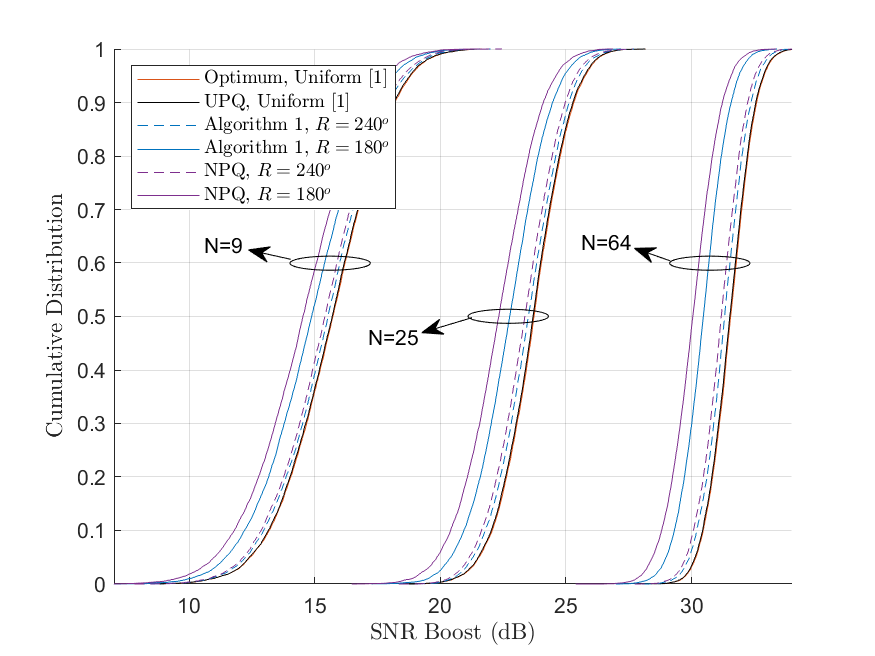}
\caption{CDF plots for SNR Boost with \cite[UPQ]{PA24}, \cite[Algorithm~1]{PA24}, Algorithm~1, and nonuniform polar quantization (NPQ), for $K=8$.}
 \label{fig:SNRBoost_CDF_K8}
\end{minipage}
\end{figure*}

\setcounter{algorithm}{0}
\begin{algorithm}[!t]
\caption{Generalized \cite[Algorithm~1]{PA24} for Nonuniform Phase Considerations}\label{alg:nonuniform_NK}
\begin{algorithmic}[1]
\State {\bf Initialization:} Compute 
$s_{nk}$ and ${\cal N} (\lambda_l)$ as in equations (\ref{eq:snk_new}) and (\ref{eq:updateRule}), respectively.
\State 
Set $\phase{\mu} = 0$. For $n=1,2,\ldots,N$, calculate and store
\[
\theta_n = \arg\max_{\theta_n\in\Phi_K} \cos(\phase{\mu} - \theta_n - \alpha_n).
\]
\State Set $g_0 = h_0 + \sum_{n=1}^N h_ne^{j\theta_n}$, ${\tt absgmax} = |g_0|$.
\For{$l = 1, 2, \ldots, L-1$}
\State For each double $\{n',k'\} \in {\cal N}(\lambda_l)$, let $\theta_{n'} = \phi_{k'}$.
\State Let
\[
g_l = g_{l-1} + \sum_{\{n',k'\}\in{\cal N}(\lambda_l)} h_{n'} \big(e^{j\theta_n} - e^{j (\phi_{k'\ominus1}) } \big)
\]
\If{$|g_l| > {\tt absgmax}$}
\State Let ${\tt absgmax} = |g_l|$
\State Store $\theta_n$ for $n=1,2,\ldots,N$
\EndIf
\EndFor
\State Read out $\theta_n^*$ as the stored $\theta_n$, $n=1,2,\ldots,N$.
\end{algorithmic}
\end{algorithm}
We present the cumulative distribution function (CDF) results for SNR Boost \cite{b1} in Fig.~\ref{fig:SNRBoost_CDF_K4} for $K=4$, and in Fig.~\ref{fig:SNRBoost_CDF_K8} for $K=8$. In these results, we consider the RIS phase range to be larger than $\pi$, i.e., $R\in\{180^\circ,240^\circ\}$, leading us to use large values of $K$ so that $R<2\pi \frac{K-1}{K}$. The CDF results are presented for $N=9,$ $25,$ and $64,$ using 10,000 realizations of the channel model defined in \cite{PA24} with $\kappa=0$. We employed uniform polar quantization (UPQ) in \cite{PA24} and the optimum algorithm \cite[Algorithm~1]{PA24} to generate the performance results for uniform discrete phase shifts and quantify the loss due to nonuniformity. We also employ Algorithm~1, and the nonuniform polar quantization (NPQ) algorithms presented in this paper, with the equally separated nonuniform discrete phase shifts structure given in Fig.~\ref{fig:phaseRange}. All algorithms ran over the same channel realization in each step. Between Fig.~\ref{fig:SNRBoost_CDF_K4} and Fig.~\ref{fig:SNRBoost_CDF_K8}, it can be seen that the loss due to the RIS phase range restriction increases for larger $K$. Note the UPQ with the uniform discrete phase shifts is always superior to NPQ, provided $R<2\pi\frac{K-1}{K}.$ However, we remark that the optimum performance provided by Algorithm~1 with nonuniform discrete phase shifts can surpass the UPQ algorithm with uniform phases. In other words, the loss due to the RIS phase range limitation is larger for the quantization approach rather than the optimum solution with $R \geq \pi$ and $\beta_n^r=1$ for $n=1,\ldots,N.$
\begin{figure*}[!t]
\centering
\begin{minipage}{0.48\textwidth}
\centering
\includegraphics[width=1.0\textwidth]{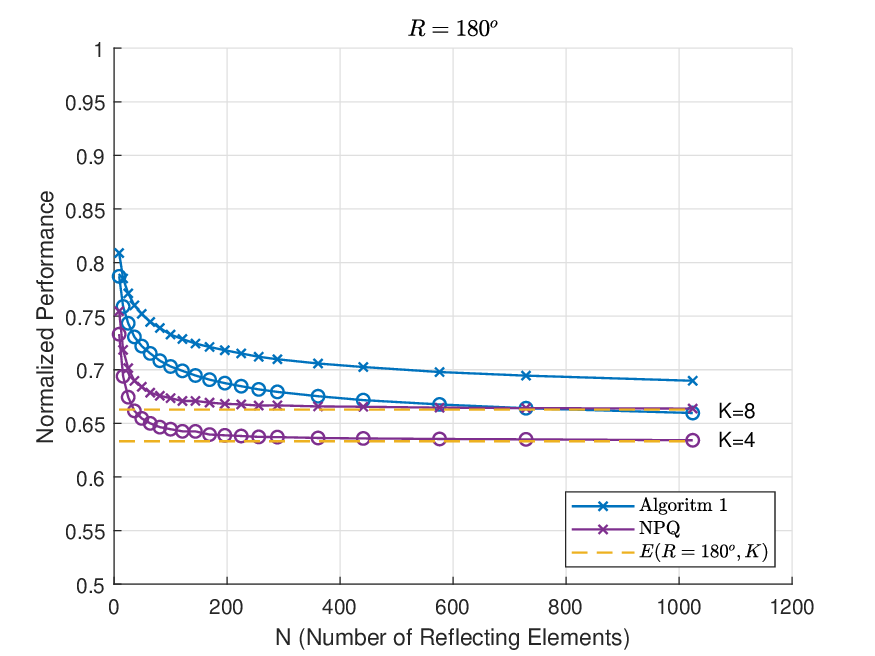}
\caption{Normalized Performance results vs. $N$, for $R=180^\circ$ and $K\in\{4,8\}$.}
 \label{fig:ApproximationPlot_R180}
\end{minipage}%
\hspace{0.03\textwidth}
\begin{minipage}{0.48\textwidth}
\centering
\includegraphics[width=1.0\textwidth]{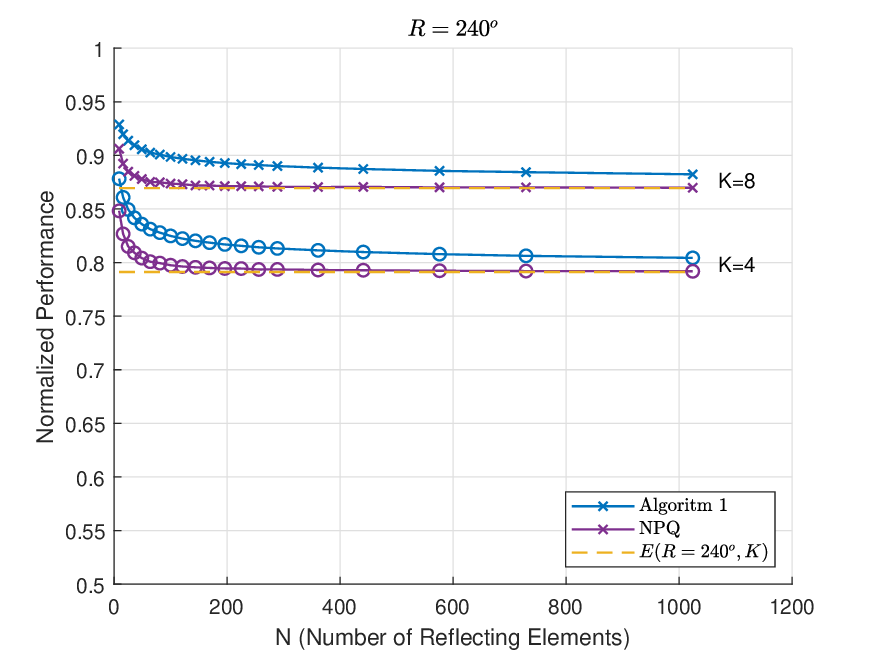}
\caption{Normalized Performance results vs. $N$, for $R=240^\circ$ and $K\in\{4,8\}$.}
 \label{fig:ApproximationPlot_R240}
\end{minipage}
\end{figure*}

Finally, the numerical results for the approximation ratio are calculated by dividing the expression $\left| \beta_0e^{j\alpha_0}+\sum_{n=1}^N \beta_n
e^{j(\alpha_n + \theta_n)}\right|^2$ to $(\sum_{n=0}^{N}\beta_n)^2$ for each channel realization and averaged. With this, the normalized performance results are presented in Fig.~\ref{fig:ApproximationPlot_R180} for $R=180^\circ$, and in Fig.~\ref{fig:ApproximationPlot_R240} for $R=240^\circ$. In both figures, the performance of NPQ converges to the approximation ratio curve for large $N$, falling in line with our analytical analysis on $E(R,K).$ Providing the optimum result, Algorithm~1 serves as an upper bound. From Fig.~\ref{fig:ApproximationPlot_R180} to Fig.~\ref{fig:ApproximationPlot_R240}, for larger $R$, the performance gap between Algorithm~1 and NPQ gets smaller. With this, we remark that increasing $R$ from $180^\circ$ to $240^\circ$ helps significantly more in terms of performance rather than increasing the number of discrete phase shifts $K$. This confirms our analysis with Fig.~\ref{fig:ERKtheoreticPlot} that the lower the RIS phase range is, the less likely it is to achieve a performance gain by increasing $K$.

Next, we will consider a larger RIS phase range limitation, i.e., $R<\pi$, and provide further analysis and insight on the received power maximization problem with RISs using nonuniform discrete phase shifts.

\subsection{Destructive Selections When $R<\pi$}
Similar to what the authors in \cite{HJA23} pointed out, we remark on an important downside of the nonuniform discrete phase shifts, especially when $R<\pi$.
We know from our Proposition~1 that the optimal phase shift selections will satisfy $\phase\mu-\theta_n-\alpha_n \in \left(-\frac{\omega_{k-1}}{2}, \frac{\omega_{k}}{2}\right)$.
So, whenever $R<\pi$, we will have an $\omega_{\Bar{k}} > \pi$ for $\Bar{k} \in \{1,2,\ldots,K\}$.
Note that there could only be one instance of $\Bar{k}$ since $\sum_{k=1}^K \omega_k = 2\pi$ must hold.
With this, depending on the optimum $\mu$, $\cos(\phase\mu - \theta_n - \alpha_n)$ can take a negative value, for some $n$.
This results in a negative contribution to the optimum $|g|$ given in (\ref{eq:resulting_g}).

\section{Global Optimum Solution With ON/OFF $\beta_n^r$}\label{sec:ONOFF}

In this section, we address the destructive selection issue by relaxing the RIS gains, i.e., $\beta_n^r \in [0,1].$ With this, we will define an updated maximization problem where we tune $\beta_n^r$ together with $\theta_n$, and develop an optimal discrete phase shift selection algorithm with ON/OFF $\beta_n^r$. We will also specify how it can converge to the optimum solution in $L \leq N(K+1)$ steps in linear time.

So far, we have developed a comprehensive analysis for the approximation ratio of nonuniform discrete phase shifts. Together with this, we provided two algorithms, i.e., NPQ and Algorithm~1, where the first is an intuitive practical algorithm and the latter achieves the global optimum with $\beta_n^r = 1$ for $n=1,\ldots, N$ in $NK$ or fewer steps, provided $R\geq\pi$. Then, we underlined the special case that arises due to the nonuniform structure of the phase shifts, or the RIS phase range constraint, that setting $\beta_n^r = 1$ for $n=1,\ldots, N$ right away can result in allowing paths that are destructive when $R<\pi$.

In this section, we will develop a new algorithm, Algorithm~2, for the special case of $R<\pi$. We will also show that this algorithm can be interchangeably used with Algorithm~1 with relaxed $\beta_n^r$. Algorithm~2 will adjust the RIS gains to manage the destructive paths through the RIS. For this purpose, we will relax the gains and redefine the optimization problem as
\begin{equation}
\begin{aligned}
 & \underset{\mbox{\boldmath$\theta$}}{\rm maximize\ } f({\mbox{\boldmath$\theta$}})\\
 & {\rm subject\ to\ } \theta_n\in \Phi_K,\ n=1, 2, \ldots, N \\
 & \quad \quad \quad \quad \,\,\,\, \beta_n^r \in [0,1],\ n=1, 2, \ldots, N
\end{aligned}
\label{eqn:problem2_relaxBeta}
\end{equation}
where
\begin{equation}
f({\mbox{\boldmath$\theta$}}) = \bigg|\beta_0e^{j\alpha_0}+\sum_{n=1}^N \beta_n \beta_n^r
e^{j(\alpha_n + \theta_n)}\bigg|^2,
\end{equation}
$\beta_n>0,\ n=0,1,\dots,N$, ${\mbox{\boldmath$\theta$}} = (\theta_1, \theta_2, \ldots, \theta_N)$, and $\alpha_n \in [-\pi,\pi)$ for $n=0,1,\dots,N$.

Similar to our {\em Lemma~1}, let
\begin{equation}\label{eq:gprime}
    g' = h_0 + \sum_{n=1}^N h_n \beta_n^{r*} e^{j\theta_n^*},
\end{equation}
so that we can state our second lemma as follows:

{\em Lemma~2:\/} To achieve the maximum of $|g'|$, a necessary condition on $(\theta_1^*, \theta_2^*, \ldots,
\theta_n^*)$ is that each $\theta_n^*$ for $n \in \{ n | \beta_n^{r*} > 0 \}$ must satisfy
\begin{equation}
\theta_n^* = \arg \max_{\theta_n\in \Phi_K} \cos(\theta_n + \alpha_n -\phase{\mu})
\label{eqn:lemma_onoff_theta}
\end{equation}
where $\phase{\mu}$ stands for the phase of optimum $\mu = g'/|g'|$ with $g'$ in equation (\ref{eq:gprime}).

{\em Proof:\/} We can rewrite equation (\ref{eq:resulting_g}) as
\begin{equation}\label{eq:resulting_gONOFF}
|g'| = \beta_0 \cos(\alpha_0 - \phase{\mu}) + \sum_{n=1}^N \beta_n \beta_n^r \cos(\theta_n^* + \alpha_n - \phase{\mu}),
\end{equation}
where $\beta_n > 0$. Therefore, for $|g'|$ to be the maximum value possible, (\ref{eqn:lemma_onoff_theta}) follows as a necessary condition, completing the proof.
\hfill$\blacksquare$

So far, similar to the development of Algorithm~1, we are proceeding with the assumption that we know the optimum $\mu$. Before coming up with the operational procedure for Algorithm~2, we will state our third lemma regarding the optimum RIS gain selection $\beta_n^{r*}$ as follows:

{\em Lemma~3:\/} Given the optimum $\mu$, the globally optimum solution will be yielded by $\beta_n^{r*} = \lceil \cos(\theta_n^* + \alpha_n - \phase{\mu}) \rceil$.

{\em Proof:\/} In equation (\ref{eq:resulting_gONOFF}), define the function $h(\beta_n^r) = \beta_n \beta_n^r \cos(\theta_n^* + \alpha_n - \phase{\mu})$ independently for every $n=1,\ldots,N$. For $|g'|$ to be the maximum value possible, given $\theta_n^*$, the function $h(\beta_n^r)$ should be maximized independently for $n=1,\ldots,N$. Note that $h(\beta_n^r)$ is a monotonic function. Therefore, to achieve the maximization in $|g'|$, $\beta_n^{r*}$ needs to satisfy
\begin{equation}\label{eq:lemmaBeta}
\beta^{r^*}_n =
\left\{
\begin{aligned}
\,\,\,1 &, \,\, \text{if}\, \cos(\theta_n^* + \alpha_n - \angle\mu) > 0, \\
\,\,\,0 &, \,\, \text{if}\, \cos(\theta_n^* + \alpha_n - \angle\mu) < 0.
\end{aligned}
\right.
\end{equation}
Therefore, without loss of generality, the optimum solution will be yielded by ON/OFF $\beta_n^r$ provided by the equality
\begin{equation}
    \beta^r_n = \lceil \cos(\theta_n + \alpha_n - \phase{\mu}) \rceil.
\end{equation}
Therefore, the proof is complete.
\hfill$\blacksquare$
\begin{figure}[!t]
	\centering
    \includegraphics[width=0.3\textwidth]{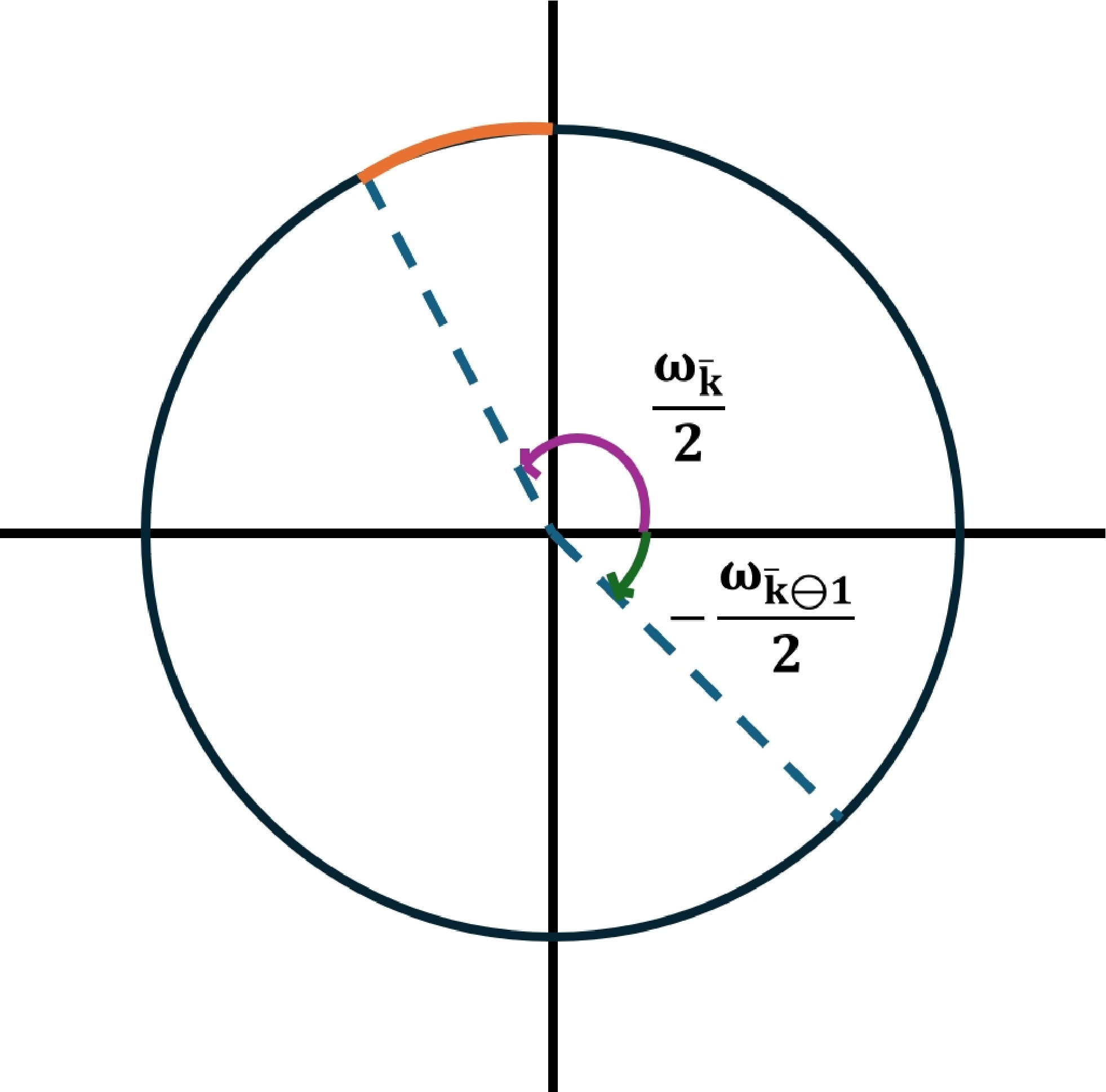}
	\caption{Range of values of $\angle\mu - \theta_n - \alpha_n$ for Case 1 with $\mu \in {\rm arc} (s_{n\Bar{k}}:s_{n,\Bar{k}\oplus1})$.}
	\label{fig:s1nk}
\end{figure}
\begin{figure}[!t]
	\centering
    \includegraphics[width=0.3\textwidth]{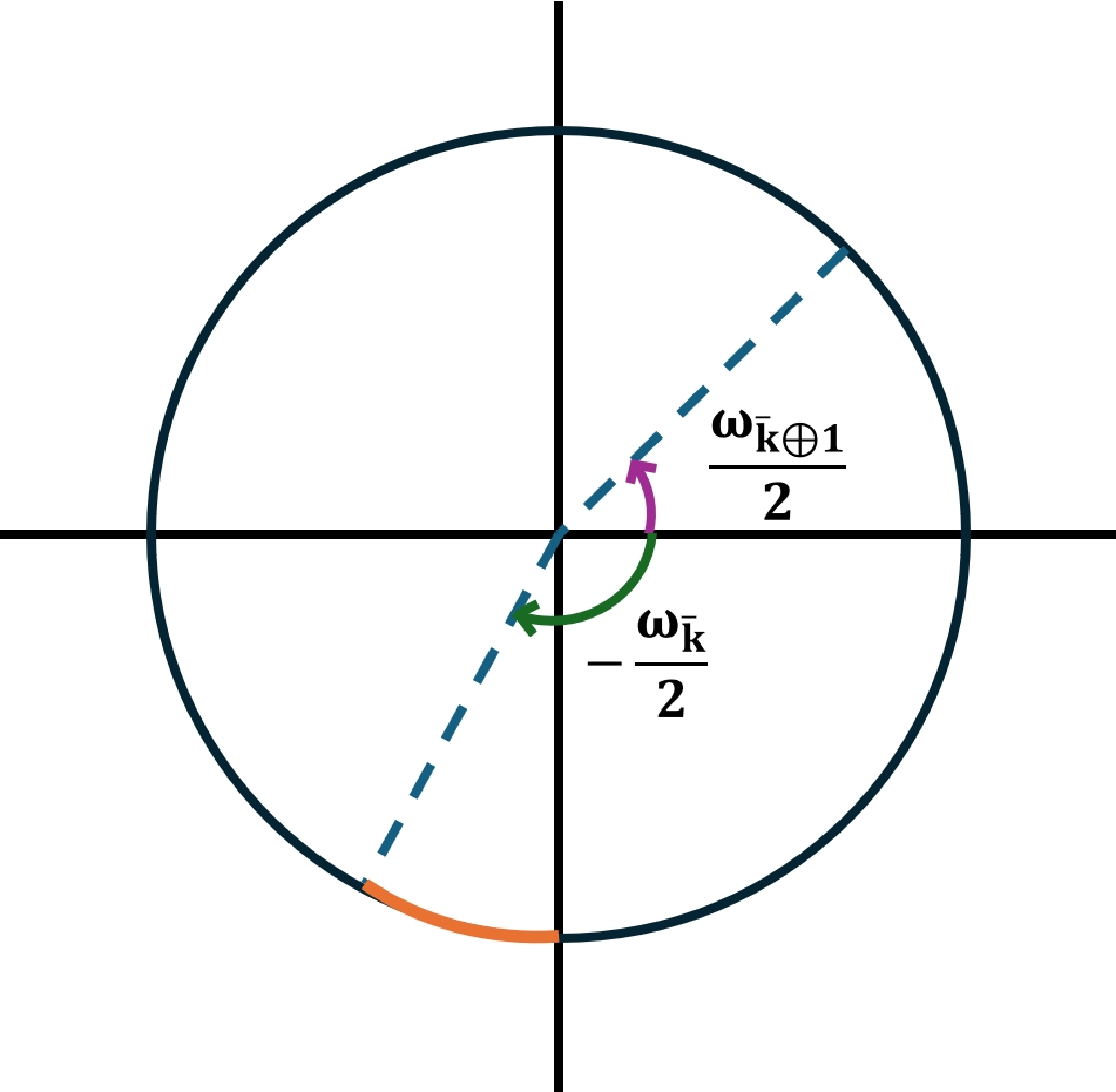}
	\caption{Range of values of $\angle\mu - \theta_n^* - \alpha_n$ for Case 2 with $\mu \in {\rm arc} (s_{n,\Bar{k}\oplus1}:s_{n,\Bar{k}\oplus2})$.}
	\label{fig:s2nk}
\end{figure}

To operate with {\em Lemma~3}, further analysis is required in terms of finding when the $\cos(\theta_n^* + \alpha_n - \angle\mu)<0$ case will arise.
For this purpose, assume for $R<\pi$ that we have the unique $\Bar{k}$ such that $\omega_{\Bar{k}}>\pi$.
We will revisit equation (\ref{eq:insideCos}) from Proposition~1, as we know from {\rm Lemma~2} that it will hold whenever $\beta_n^r>0$.
Our $\theta_n^*$ selections will make sure that $\phase\mu-\theta_n-\alpha_n \in \left(-\frac{\omega_{k-1}}{2}, \frac{\omega_{k}}{2}\right)$, given that $\mu \in {\rm arc} (s_{nk}:s_{n,k+1}).$
Consider two cases, $\mu \in {\rm arc} (s_{n\Bar{k}}:s_{n,\Bar{k}\oplus1})$ and $\mu \in {\rm arc} (s_{n,\Bar{k}\oplus1}:s_{n,\Bar{k}\oplus2})$.
As shown in Fig. \ref{fig:s1nk} and Fig. \ref{fig:s2nk}, in both cases the cosine value in (\ref{eq:resulting_gONOFF}) can take a negative value, i.e., $\cos(\theta_n + \alpha_n - \angle\mu) < 0$, resulting in the selection of $\beta_n^r=0$.
With this observation, we propose the following proposition to be able to operate with {\em Lemma~2} and {\em Lemma~3}.

{\em Proposition 2:\/} Let $s^1_{n,\Bar{k}\oplus1} = e^{j(\alpha_n + \phi_{\Bar{k}} + \frac{\pi}{2})}$ and $s^2_{n,\Bar{k}\oplus1} = e^{j(\alpha_n + \phi_{\Bar{k}\oplus1} - \frac{\pi}{2})}$. A sufficient condition for $\beta_n^{r*}=0$ is
\begin{equation}
    \mu \in {\rm arc} (s^1_{n,\Bar{k}\oplus1} : s^2_{n,\Bar{k}\oplus1}).
\label{eqn:sufficiencyBeta}
\end{equation}

{\em Proof:\/} Consider the two and only cases that $\cos(\theta_n + \alpha_n - \angle\mu)$ can take a negative value.

First, assume $\mu \in {\rm arc} (s_{n\Bar{k}}:s_{n,\Bar{k}\oplus1})$. Since $\omega_{\Bar{k}}/2>\pi/2$, the cosine value can take a negative value as shown in Fig. \ref{fig:s1nk}. This happens if $\mu \in {\rm arc} (s_{n,\Bar{k}\oplus1} e^{-j(\omega_{\Bar{k}}-\pi)/2} : s_{n,\Bar{k}\oplus1})$, as there is no $\theta_n \in \Phi_K$ such that $\cos(\theta_n + \alpha_n - \angle\mu)>0.$

Second, assume $\mu \in {\rm arc} (s_{n,\Bar{k}\oplus1}:s_{n,\Bar{k}\oplus2})$. Since $-\omega_{\Bar{k}}/2<-\pi/2$, the cosine value can take a negative value as shown in Fig. \ref{fig:s2nk}. This happens if $\mu \in {\rm arc} (s_{n,\Bar{k}\oplus1} : s_{n,\Bar{k}\oplus1} e^{j(\omega_{\Bar{k}}-\pi)/2})$, as there is no $\theta_n \in \Phi_K$ such that $\cos(\theta_n + \alpha_n - \angle\mu)>0.$

Finally, the two cases together can be expressed as a single arc around $s_{n,\Bar{k}\oplus1}$ by using $s^1_{n,\Bar{k}\oplus1}$ and $s^2_{n,\Bar{k}\oplus1}$ as $\mu \in {\rm arc} (s_{n,\Bar{k}\oplus1} e^{-j(\omega_{\Bar{k}}-\pi)/2} : s_{n,\Bar{k}\oplus1} e^{j(\omega_{\Bar{k}}-\pi)/2})$. Since $\phi_{\Bar{k}\oplus1}=\phi_{\Bar{k}} + \omega_{\Bar{k}}$, the same arc can be expressed as $\mu \in {\rm arc} (e^{j(\alpha_n + \phi_{\Bar{k}} + \frac{\pi}{2})} : e^{j(\alpha_n + \phi_{\Bar{k}\oplus1} - \frac{\pi}{2})})$. Thus, the proof is complete.
\hfill$\blacksquare$

With Proposition~1 and Proposition~2 together, we need to consider $K+1$ arcs that the optimum $\mu$ can be in for every $n=1,\ldots,N$ independently as there is an extra arc introduced in Proposition~2 for $n=1,\ldots,N.$ This is because, when $R<\pi$ and $\omega_{\Bar{k}}>\pi$, we will let $s_{n,\Bar{k}\oplus1} = \{s^1_{n,\Bar{k}\oplus1},s^2_{n,\Bar{k}\oplus1}\}$ so that $s_{n,\Bar{k}\oplus1}$ will encode two complex numbers.
\begin{algorithm}[!t]
\caption{Extended Algorithm~1 for the Special Condition When $R<\pi$.}
\begin{algorithmic}[1]
\State {\bf Initialization:} Compute 
$s_{nk}$ and ${\cal N} (\lambda_l)$ as in Proposition~2 and equation (\ref{eq:updateRule}), respectively.
\State 
Set $\phase{\mu} = 0$. For $n=1,2,\ldots,N$, calculate
\[
\theta_n = \arg\max_{\theta_n\in\Phi_K} \cos(\phase{\mu} - \theta_n - \alpha_n).
\]
\State Set $\beta_n^r = \left\lceil \cos(\phase{\mu} - \theta_n - \alpha_n) \right \rceil$ for $n=1,2,\dots,N$.
\State Update $\theta_n = \phi_{\Bar{k} \oplus 1}$ for $n \in \{ n | \beta_n^r=0\}$, and store $\theta_n, \forall n$.
\State Set $g_0 = h_0 + \sum_{n=1}^N h_n \beta_n^r e^{j\theta_n}$, ${\tt absgmax} = |g_0|.$
\For{$l = 1, 2, \ldots, L'-1$}
\State Set $g_{{\tt update}}=0.$
\For{each double $\{n',k'\} \in {\cal N}(\lambda_l)$}
\If{$\beta_{n'}=1$}
    \If{$k' = \Bar{k}\oplus1$}
        \State Set $\beta_{n'}=0$ and $\theta_{n'}=\phi_{k'}$
        \State Let
        \[
        g_{{\tt update}} - h_{n'} e^{j \phi_{\Bar{k}}} \leftarrow g_{{\tt update}}.
        \]
    \Else
        \State Set $\theta_{n'}=\phi_{k'}$
        \State Let
        \[
        g_{{\tt update}} + h_{n'} \big(e^{j\theta_{n'}} - e^{j (\phi_{k'\ominus1}) } \big)\leftarrow g_{{\tt update}}.
        \]
    \EndIf
\Else
    \State Set $\beta_{n'}=1$
    \State Let
    \[
    g_{{\tt update}} + h_{n'} e^{j\theta_{n'}}\leftarrow g_{{\tt update}}.
    \]
\EndIf
\EndFor
\State Let $g_l = g_{l-1} + g_{{\tt update}}$
\If{$|g_l| > {\tt absgmax}$}
\State Let ${\tt absgmax} = |g_l|$
\State Store $\beta_n^r$ and $\theta_n$ for $n=1,2,\ldots,N$
\EndIf
\EndFor
\State Read out $\beta_n^{r^*}$ and $\theta_n^*$, $n=1,2,\ldots,N$.
\end{algorithmic}
\end{algorithm}

Similar to Algorithm~1, after eliminating the duplicates among $s_{nk}$, in total we will consider $L' \leq N(K+1)$ arcs leading to $\mathcal{O}(N(K+1))$ complexity at maximum. To achieve the linear time complexity while running the algorithms, similar to the ideas in \cite{b1,PA24}, we present the approach given in this section under Algorithm~2, such that in each of the search steps, only one or a small number of elements are updated.
\begin{figure*}[!t]
\centering
\begin{minipage}{0.48\textwidth}
\centering
\includegraphics[width=1.1\textwidth]{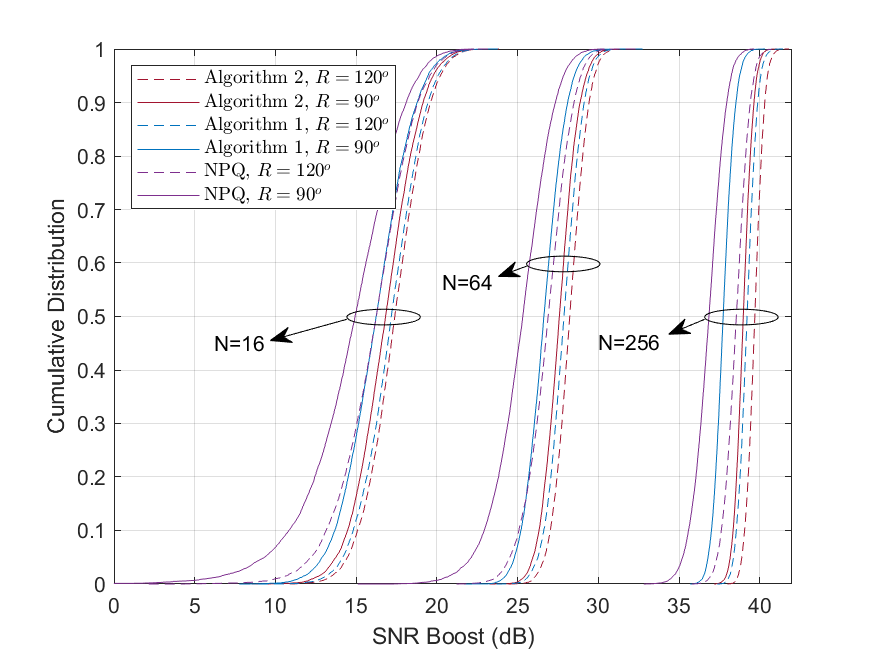}
\caption{CDF plots for SNR Boost with nonuniform polar quantization (NPQ), Algorithm~1, and Algorithm~2 for $K=2$ and $R\in\{90^\circ,120^\circ\}$.}
 \label{fig:SNRBoost_Algo2_CDF_K2}
\end{minipage}%
\hspace{0.03\textwidth}
\begin{minipage}{0.48\textwidth}
\centering
\includegraphics[width=1.1\textwidth]{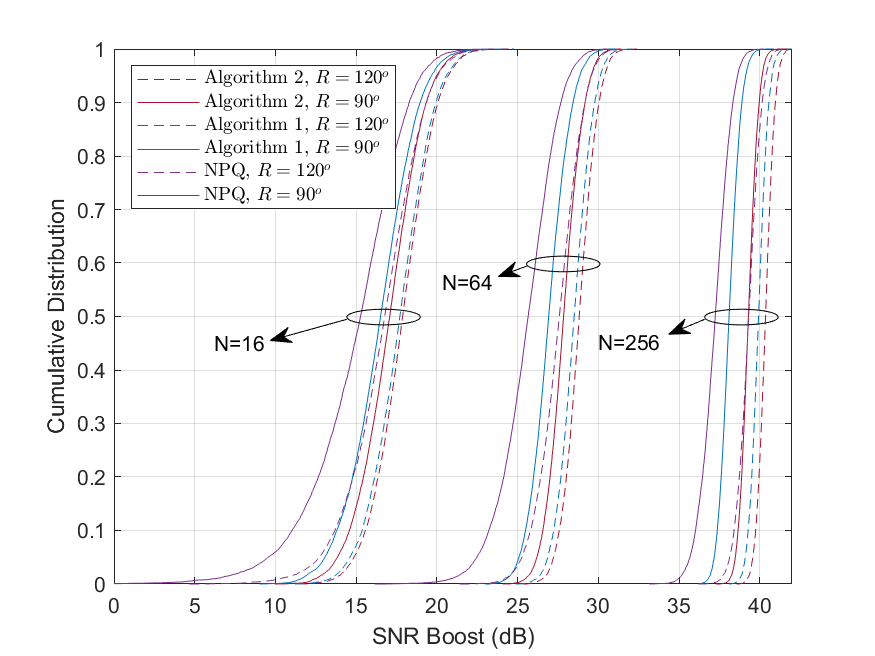}
\caption{CDF plots for SNR Boost with nonuniform polar quantization (NPQ), Algorithm~1, and Algorithm~2 for $K=4$ and $R\in\{90^\circ,120^\circ\}$.}
 \label{fig:SNRBoost_Algo2_CDF_K4}
\end{minipage}
\end{figure*}

We present the cumulative distribution function (CDF) results for SNR Boost in Fig.~\ref{fig:SNRBoost_Algo2_CDF_K2} for $K=2$, and in Fig.~\ref{fig:SNRBoost_Algo2_CDF_K4} for $K=4$. In these results, we consider a notable limitation on the RIS phase range such that $R < \pi$, i.e., $R\in\{90^\circ,120^\circ\}$. The CDF results are presented for $N=16,$ $64,$ and $256,$ using 10,000 realizations of the channel model defined in \cite{PA24} with $\kappa=0$. The discrete phase shift selections are equally separated and chosen as given in Fig.~\ref{fig:phaseRange}. We employed Algorithm~1, Algorithm~2, and NPQ algorithms that we proposed in this paper. Since we have $R<\pi$, Algorithm~1 will only serve as a pseudo-optimal solution, assuming that $\beta_n^r$ are strictly $1$ for all $n$, so that we can observe the effect of destructive paths and ON/OFF keying. All algorithms ran over the same channel realization in each step. It can be seen that the gap between Algorithm~2 and the other algorithms increases for larger $N$, as well as for smaller $R$. This signifies the power of using ON/OFF $\beta_n^r$ with larger RISs, having more phase range limitations. Furthermore, using $K=4$ instead of $K=2$ mostly impacts the performance of NPQ with $R=120^\circ$, making it more desirable due to its low complexity.
\begin{figure*}[!t]
\centering
\begin{minipage}{0.48\textwidth}
\centering
\includegraphics[width=1.0\textwidth]{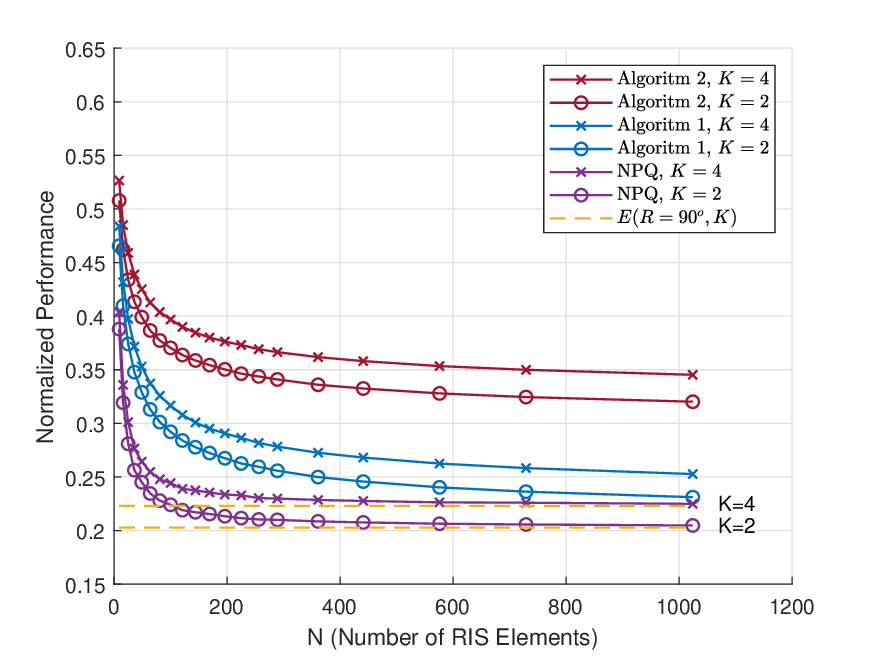}
\caption{Normalized Performance results vs. $N$, for $R=90^\circ$ and $K\in\{2,4\}$.}
 \label{fig:ApproximationPlot_R90}
\end{minipage}%
\hspace{0.03\textwidth}
\begin{minipage}{0.48\textwidth}
\centering
\includegraphics[width=1.0\textwidth]{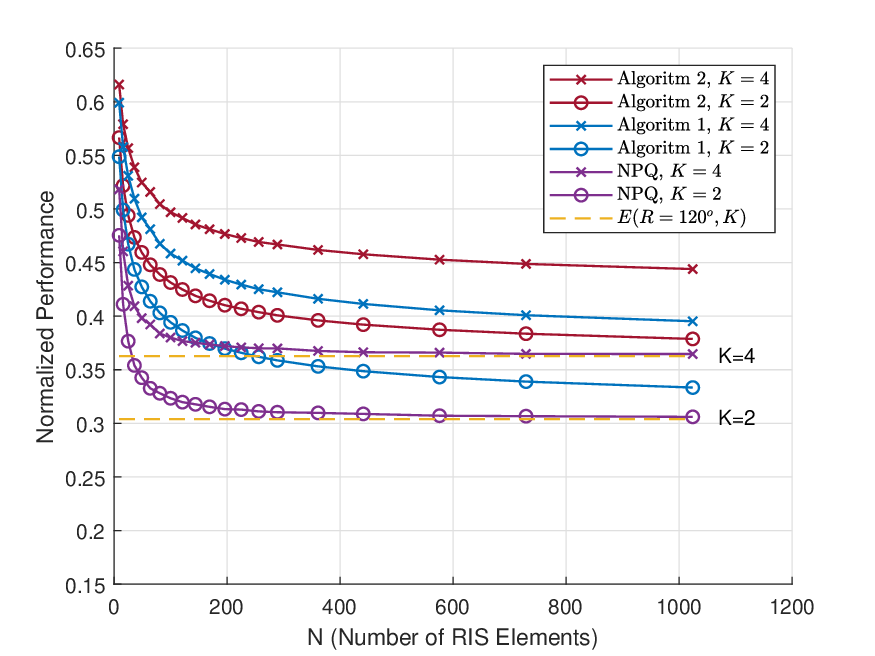}
\caption{Normalized Performance results vs. $N$, for $R=120^\circ$ and $K\in\{2,4\}$.}
 \label{fig:ApproximationPlot_R120}
\end{minipage}
\end{figure*}

With Algorithm~2 and $R<\pi$, the normalized performance results are presented in Fig.~\ref{fig:ApproximationPlot_R90} for $R=90^\circ$, and in Fig.~\ref{fig:ApproximationPlot_R120} for $R=120^\circ$. In both figures, the performance of NPQ converges to the approximation ratio curve for large $N$, again confirming our analytical analysis on $E(R,K).$ Similar to the CDF plots, the performance gain from using Algorithm~2 over both NPQ and Algorithm~1 increases for larger $N$. Also, if $R$ is sufficiently low, Algorithm~2 is always superior to Algorithm~1. Similarly, Algorithm~1 is always superior to NPQ, even if a larger $K$ is used for the latter. The underlying reason for this again is that the performance gain from using larger $K$ diminishes significantly for low $R$.

\begin{figure*}[!t]
\centering
\begin{minipage}{0.48\textwidth}
\centering
\includegraphics[width=1.0\textwidth]{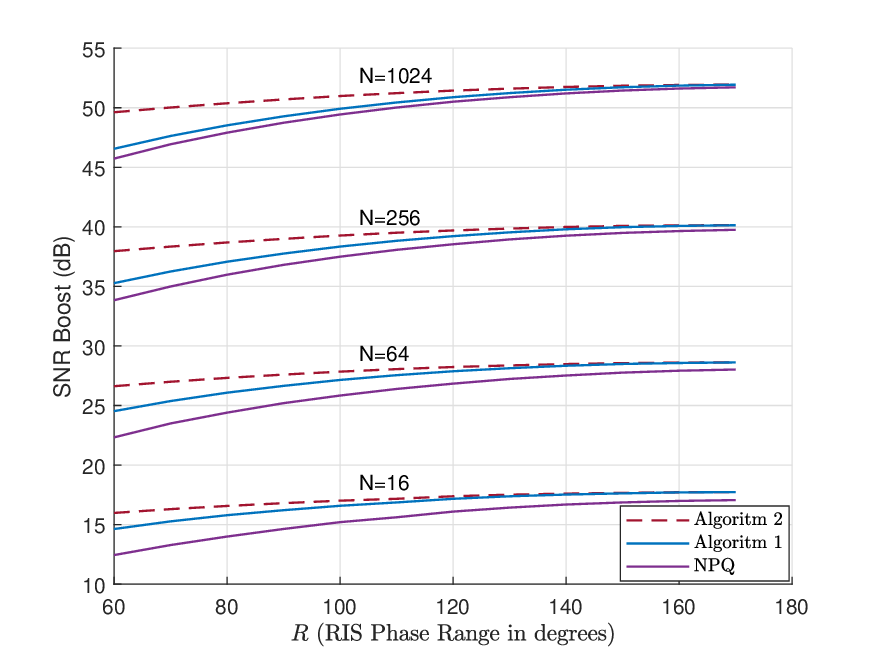}
\caption{Average SNR Boost vs. $R$, for $K=2$ and $N\in\{16,64,256,1024\}$.}
 \label{fig:SNRBoost_vs_R_K2}
\end{minipage}%
\hspace{0.03\textwidth}
\begin{minipage}{0.48\textwidth}
\centering
\includegraphics[width=1.0\textwidth]{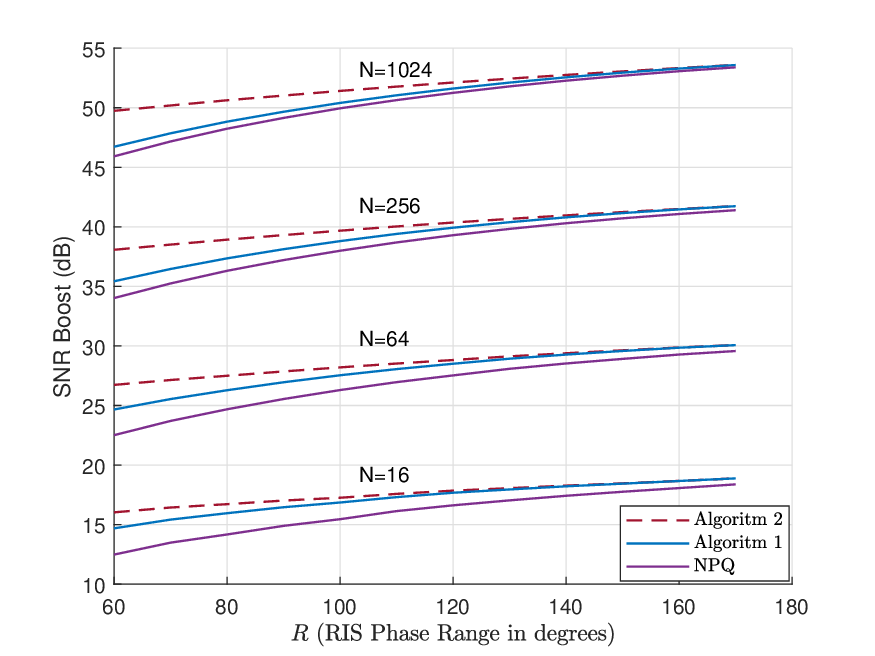}
\caption{Average SNR Boost vs. $R$, for $K=4$ and $N\in\{16,64,256,1024\}$.}
 \label{fig:SNRBoost_vs_R_K4}
\end{minipage}
\end{figure*}

Finally, we present the average SNR Boost results of our proposed algorithms versus $R$ in Fig.~\ref{fig:SNRBoost_vs_R_K2} for $K=2$, and in Fig.~\ref{fig:SNRBoost_vs_R_K4} for $K=4$. Both figures show that the average performance of Algorithm~1 converges to that of Algorithm~2, as $R$ approaches $\pi$. On the other hand, NPQ can provide an average SNR Boost that is significantly close to both Algorithm~1 and Algorithm~2 as $R$ increases, for large $N$. Both Fig.~\ref{fig:SNRBoost_vs_R_K2} and Fig.~\ref{fig:SNRBoost_vs_R_K4} suggest in a sense that Algorithm~1 and Algorithm~2 can be used interchangeably to solve the problem in (\ref{eqn:problem2_relaxBeta}), where the selection depends on whether $R<\pi$ or $R\geq \pi$.

\subsection{Generalized Algorithm~1 With Relaxed $\beta_n^r$}
Finally, we remark that the development of Algorithm~2 follows from the strict limitation on the RIS phase range, i.e., $R<\pi$. Otherwise, an important side conclusion that follows from {\em Lemma~3\/} is that, Algorithm~1 can be extended to solve the problem in (\ref{eqn:problem2_relaxBeta}) with $\beta_n^r \in [0,1]$ for $n=1,\ldots,N$, when $R>\pi$. With $R>\pi$, we know from {\em Lemma~3\/} that the solution that yields the global optimum will select $\beta_n^r = 1$ for $n=1,\ldots,N$. Therefore, both Algorithm~1 and Algorithm~2 can be used to solve the general problem in (\ref{eqn:problem2_relaxBeta}) for $R \geq \pi$ and $R<\pi$, respectively. With this, the number of required steps in the for loop would reduce from $N(K+1)$ to $NK$. Further analysis regarding the number of required steps and complexity is provided in the following section.

\section{Convergence to Optimality and Complexity}
We will now discuss the convergence of Algorithm~1 and Algorithm~2 to the optimal solution for $\beta_n^r=1$ and $\beta_n^r \in [0,1]$, $n=1,\ldots,N$, respectively. We know from {\em Lemma~1\/} and {\em Proposition 1\/} that Algorithm~1 will converge to the global optimum. Whereas, convergence to the global optimality of Algorithm~2 is guaranteed by {\em Lemma 1\/}, {\em Lemma 2\/}, and {\em Proposition 1\/}. Next, we will discuss the required complexity of both algorithms to achieve global optimality.

First, for Algorithm~1, the for loop from Step 4 to Step 11 takes $\sum_{l=1}^{L}\mathcal{O}(|{\cal N}(\lambda_l)|) = \mathcal{O}(NK)$ steps. With this, two vector additions are performed for each updated element. Together with the $N$ vector additions in Step 3, Algorithm~1 incurs $N(2K+1)$ vector additions in total.

Second, for Algorithm~2, the for loop from Step 6 to Step 27 takes $\sum_{l=1}^{L'}\mathcal{O}(|{\cal N}(\lambda_l)|) = \mathcal{O}(N(K+1))$ steps. With this, there are $K+1$ arcs to be considered for each element, where only one vector addition is performed for two of those arcs and two vector additions are performed for the remaining $K-1$ arcs. Therefore, for each element, $2\times1 + (K-1)\times2=2K$ vector additions are performed. With the $N$ vector additions in Step 5, this amounts to $N(2K+1)$ vector additions in total. Note that, although the number of steps is larger for Algorithm~2, the total number of vector additions performed is the same as for Algorithm~1.

Finally, similar to \cite{b1}, assuming $\alpha_n$ are uniformly distributed, it can be argued that the sorting in ${\cal N}(\lambda_l)$ will take $\mathcal{O}(N)$ time on average. Therefore, since in each step of both Algorithm~1 and Algorithm~2 only one or a small number of elements are updated, the time complexity of both algorithms will be linear in $N$. Also, our proposed algorithms converge to the global optimum by performing $N(2K+1)$ vector additions in total, whereas, in \cite{HJA23}, the proposed solution is claimed to achieve the optimum with $N(2K+3)$ vector additions.
\begin{figure*}[!t]
\centering
\begin{minipage}{0.48\textwidth}
\centering
\includegraphics[width=1.0\textwidth]{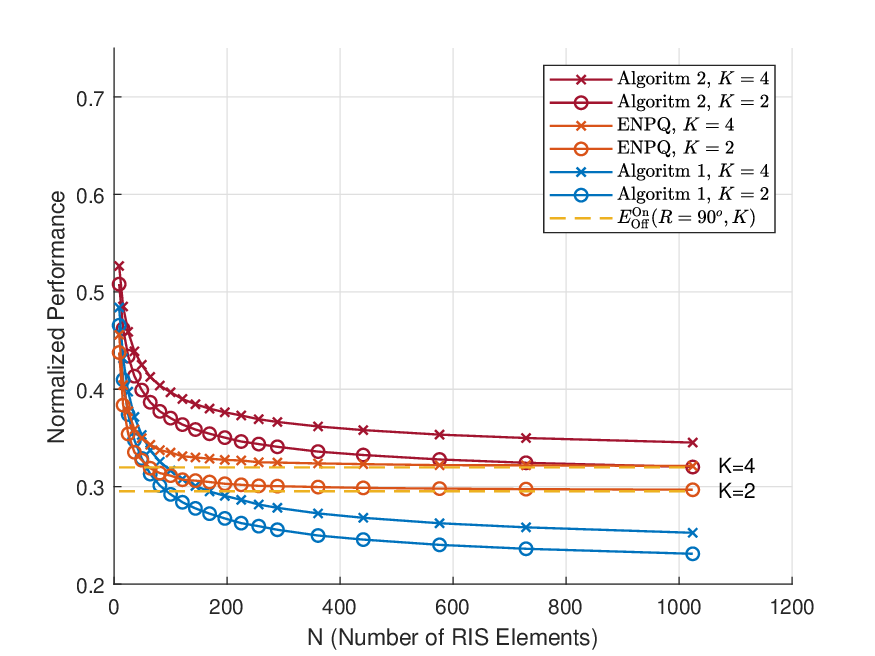}
\caption{Normalized Performance results vs. $N$, for $R=90^\circ$ and $K\in\{2,4\}$.}
 \label{fig:ApproximationPlot_OnOff_R90}
\end{minipage}%
\hspace{0.03\textwidth}
\begin{minipage}{0.48\textwidth}
\centering
\includegraphics[width=1.0\textwidth]{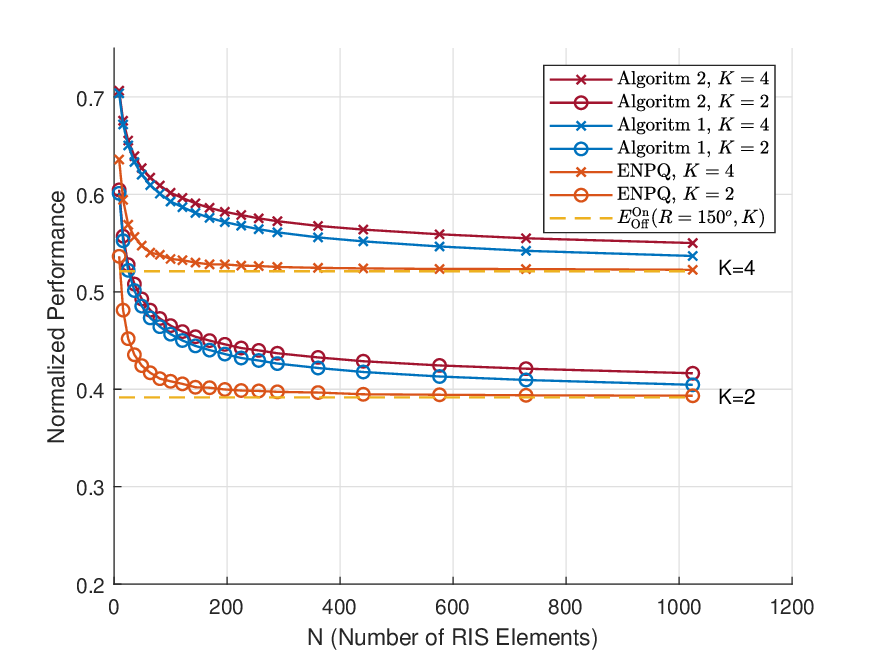}
\caption{Normalized Performance results vs. $N$, for $R=150^\circ$ and $K\in\{2,4\}$.}
 \label{fig:ApproximationPlot_OnOff_R150}
\end{minipage}
\end{figure*}

\section{Revisiting the Quantization Solution With ON/OFF $\beta_n^r$: Extended Nonuniform Polar Quantization}\label{ch:quantization_ONOFF}
In this section, we will propose a novel quantization algorithm by enhancing the NPQ algorithm with ON/OFF $\beta_n^r$ selections. The importance of the ON/OFF selections has been established so far, showing significant performance gains for $R<\pi$. A similar approach to exploit $\beta_n^r$ in Algorithm~2 can be used for the quantization solution.
\begin{figure}[!t]
	\centering
\includegraphics[width=0.48\textwidth]{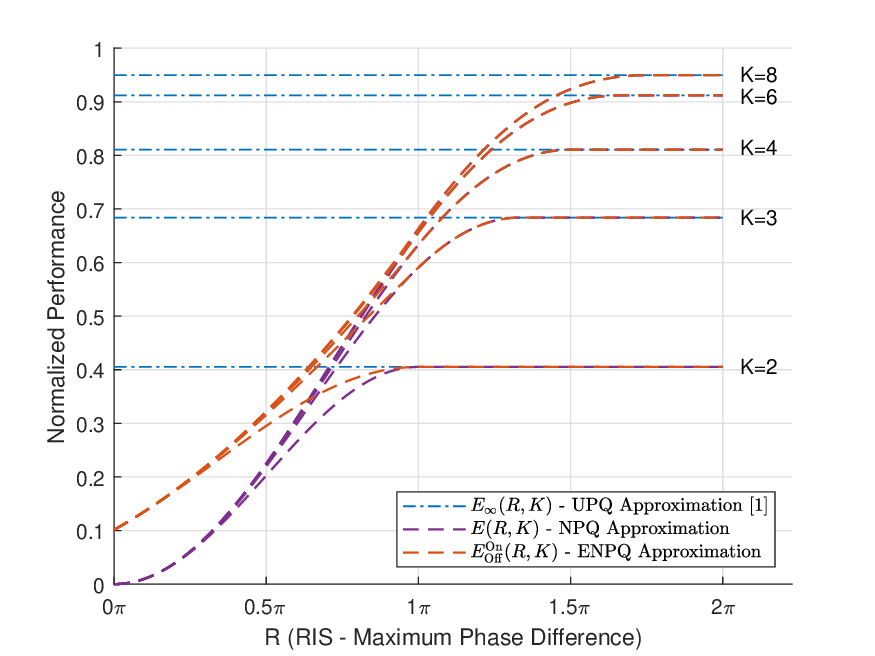}
	\caption{$E^{\text{on}}_{\text{off}}(R,K)$ vs $R$ for $K \in \{2,3,4,6,8\}$}
	\label{fig:ERK_onoff_theoreticPlot}
\end{figure}

The quantization approach comes from selecting the closest option from the phase shifts set to the continuous solution, which can achieve the maximum possible received power given by $(\sum_{n=0}^N \beta_n)^2$. Similar to our analysis in Section~\ref{sec:ONOFF}, let $\delta_n = \theta_n^{\text{NPQ}} - \theta_n^{\text{cont}}$ for $n=1,\ldots,N$. When $R<\pi$, depending on the value $\theta_n^{\text{cont}}$, the difference between $\theta_n^{\text{NPQ}}$ and $\theta_n^{\text{cont}}$ in (\ref{eqn:thetaCont}) can be greater than $\frac{\pi}{2}$, or less than $-\frac{\pi}{2}$, i.e., $|\delta_n| > \frac{\pi}{2}$. Therefore, such a path through the $n$-th RIS element would contribute destructively to the overall performance, as could be deduced from $E(R,K) = (\mathbb{E}[\cos(\delta_n)])^2$. With the adjustable RIS gains, this can be eliminated by an OFF selection, i.e., $\beta_n^r = 0$. Therefore, we define the extended nonuniform polar quantization (ENPQ) algorithm with ON/OFF $\beta_n^r$, which is an algorithm to select the RIS coefficients, as 
\begin{equation}\label{eq:ENPQ}
    {\rm w}_n^{\text{ENPQ}} = \lceil \cos(\delta_n) \rceil \exp\left({j \theta_n^{\text{NPQ}}}\right),
\end{equation}
where $\theta_n^{\text{NPQ}}$ are selected by the NPQ algorithm, and $\delta_n = \theta_n^{\text{NPQ}} - \theta_n^{\text{cont}}$. Note that for $R \geq \pi$, ENPQ will select the same RIS coefficients as the NPQ algorithm, because $|\delta_n| > \frac{\pi}{2}$ will never occur.

\section{Approximation Ratio Calculation for ENPQ}
We extend our approximation ratio calculations to find the approximation ratio for the ENPQ algorithm, i.e., $E^{\text{on}}_{\text{off}}(R,K)$. With the independence assumption among $\delta_n$, it can be deduced from (\ref{eqn:frxApxCos})--(\ref{eq:approxCosine}) that $E^{\text{on}}_{\text{off}}(R,K) = \left(\mathbb{E}\left[\lceil\cos(\delta_n)\rceil\cos\left(\delta_n\right)\right]\right)^2 + \left(\mathbb{E}\left[\lceil\cos(\delta_n)\rceil\sin\left(\delta_n\right)\right]\right)^2$ by including the $\lceil\cos(\delta_k)\rceil$ and $\lceil\cos(\delta_l)\rceil$ terms. Due to the symmetry in $\delta_n$, $\left(\mathbb{E}\left[\lceil\cos(\delta_n)\rceil\sin\left(\delta_n\right)\right]\right)^2=0$, so that $E^{\text{on}}_{\text{off}}(R,K) = \left(\mathbb{E}\left[\lceil\cos(\delta_n)\rceil\cos\left(\delta_n\right)\right]\right)^2$. Now, with the PDF of $\delta_n$ given in Fig.~\ref{fig:PDFdeltan}, the expected value can be calculated as follows:
\begin{equation}\label{eq:approxSincu_OnOff1}
\mathbb{E} \left[\lceil\cos(\delta_n)\rceil\cos\left(\delta_n\right)\right]
= 2 \bigg[ \int_{0}^{\frac{R}{2(K-1)}} \lceil\cos(\delta_n)\rceil\cos(\delta_n) \frac{K}{2\pi}\,d\delta_n + \int_{\frac{R}{2(K-1)}}^{\pi-R/2} \lceil\cos(\delta_n)\rceil\cos(\delta_n) \frac{1}{2\pi}\,d\delta_n \bigg]
\end{equation}
where in the first integral, $\lceil\cos(\delta_n)\rceil=1$ as $\frac{R}{2(K-1)} < \frac{\pi}{2}$. Whereas, in the second integral, when $\pi-R/2 > \frac{\pi}{2}$, i.e., $R<\pi$, the upper limit of the integral should be updated as $\frac{\pi}{2}$ as $\lceil\cos(\delta_n)\rceil=0$ when $|\delta_n|>\frac{\pi}{2}$. Therefore, (\ref{eq:approxSincu_OnOff1}) is rewritten as
\begin{align}
    \mathbb{E} \left[\lceil\cos(\delta_n)\rceil\cos\left(\delta_n\right)\right] & = \mathbb{E} \left[\cos\left(\delta_n\right)\right] \nonumber \\
    &= 2 \left[\int_{0}^{\frac{R}{2(K-1)}} \cos(\delta_n) \frac{K}{2\pi}\,d\delta_n + \int_{\frac{R}{2(K-1)}}^{\frac{\pi}{2}} \cos(\delta_n) \frac{1}{2\pi}\,d\delta_n\right] \nonumber \\
    &= \frac{1}{\pi}\left[(K-1)\sin\left(\frac{R}{2(K-1)}\right) + 1\right]. \label{eq:approxSin_onoff}
\end{align}
where we keep the $\sin(\cdot)$ function instead of $\sinc(\cdot)$ this time for a clearer notation. Thus, the approximation ratio for the ENPQ algorithm is
\begin{equation}
    E^{\text{on}}_{\text{off}}(R,K) = \frac{1}{\pi^2}\left[(K-1)\sin\left(\frac{R}{2\pi(K-1)}\right) + 1\right]^2.
\end{equation}
An illustration for the theoretical calculations of the approximation $E^{\text{on}}_{\text{off}}(R,K)$ is given in Fig. \ref{fig:ERK_onoff_theoreticPlot}, where it can be seen that $E^{\text{on}}_{\text{off}}(R,K)$ converges to the approximation ratio of the NPQ, i.e., $E(R,K)$, as $R$ reaches $\pi$.
We remark on the importance of using the ON/OFF $\beta_n^r$ for $R<\pi$. This can be seen from Fig. \ref{fig:ERK_onoff_theoreticPlot} that as $R$ approaches zero, while $E(R,K)$ becomes zero with all the elements being ON, $E^{\text{on}}_{\text{off}}(R,K)$ on the other hand becomes $0.1$. Therefore, even with $\theta_n$ being the same for $n=1,\dots,N$, i.e., no phase shifts selection as $R$ becomes zero, ON/OFF selections solely could beat the performance of $\beta_n^r$ for up to $K=8$ phase shift selections when $R<60^o$.
Furthermore, when there are $K=2$ discrete phase shifts with ON/OFF $\beta_n^r$, the average performance is better than the case when $\beta_n^r=1$ with up to $K=8$ discrete phase shifts, for $R<115^\circ$.

With NPQ and $R<\pi$, the normalized performance results are presented in Fig.~\ref{fig:ApproximationPlot_OnOff_R90} for $R=90^\circ$, and in Fig.~\ref{fig:ApproximationPlot_OnOff_R150} for $R=150^\circ$. As a validity check for our $E^{\text{on}}_{\text{off}}(R,K)$ calculation, we remark that the numerical results for ENPQ indeed converge to the theoretical approximation ratio.
For a lower value of $R=90^\circ$ in Fig.~\ref{fig:ApproximationPlot_OnOff_R90}, the simple quantization approach with ON/OFF $\beta_n^r$ selections outperforms the optimum solution with $\beta_n^r=1$ for $N\geq100$. On the other hand, when $R$ is high enough, say $R=150^\circ$ as in Fig.~\ref{fig:ApproximationPlot_OnOff_R150}, there is not such a loss due to the limited RIS phase range that ENPQ could exploit with ON/OFF $\beta_n^r$, resulting in Algorithm~1 being superior.

\section{Conclusion}
In this paper, to maximize the received power at a UE, we provided necessary and sufficient conditions for determination of the RIS coefficients that are subject to nonuniform discrete phase shifts. Also, we established a foundation on the RIS phase range $R$ with the nonuniform discrete phase shifts structure. We proved that the optimum placement of the nonuniform discrete phase shifts would be equally separated over the RIS phase range. Then, we showed that whenever $R<\pi$, adjusting RIS gains can bring significant performance, and surprisingly, the globally optimum solution would be yielded by these adjustable gains being either one or zero, i.e., the RIS elements being either ON or OFF.

We employed the necessary and sufficient conditions to achieve the optimum solution in $NK$ or fewer steps when $R\geq \pi$ and $N(K+1)$ or fewer steps when $R<\pi$, where in both cases, there are only $N(2K+1)$ complex vector additions are performed. Therefore, the globally optimum solution can be achieved in linear time.

In addition to the optimum algorithms, we also calculated the approximation ratio for the nonuniform discrete phase shifts by employing the intuitive quantization algorithm. Furthermore, with the ON/OFF RIS gains, we proposed a novel quantization algorithm named ENPQ, a low-complexity algorithm that can bring significant performance when there is a notable limitation in the RIS phase range, with which we also provided a secondary closed-form solution for the approximation ratio for nonuniform discrete phase shifts. With both our theoretical and simulation results, we showed that using more than $K=3$ discrete phase shifts and more than $K=4$ discrete phase shifts brings negligible performance, when $R<\pi/2$ and $R<\pi$, respectively. Furthermore, we showed that when $R<2\pi/3$, ON/OFF selections for the RIS elements can bring significant performance compared to the scenario when the RIS gains are strictly one.

Finally, we remark that the intuitive quantization algorithms are observed to achieve close-to-optimum performances under various scenarios, making them a powerful option as they are low-complexity algorithms. Especially, when $R<\pi$, the novel quantization algorithm proposed in this paper, i.e., ENPQ, can provide significant performance gains compared to NPQ and Algorithm~1 by exploiting the ON/OFF selections.

%
%

\bibliographystyle{IEEEtran}
\bibliography{ref}
\end{document}